# Gas-regulation of galaxies: the evolution of the cosmic sSFR, the metallicity-mass-SFR relation and the stellar content of haloes

Simon J. Lilly[1]

C. Marcella Carollo[1]

Antonio Pipino[1]

Alvio Renzini[2,3]

Yingjie Peng[1]

1.  Institute for Astronomy, Department of Physics, ETH Zurich, 8093 Zurich, Switzerland.

2.  Department of Physics and Astronomy Galileo Galilei, Universita degli Studi di Padova, via Marzolo 8, I-35131 Padova, Italy.

3.  INAF – Osservatorio Astronomico di Padova, vicolo dell'Osservatorio 5, I-35122 Padova, Italy

**Abstract**

A very simple physical model of galaxies, in which the formation of stars is instantaneously regulated by the mass of gas in a reservoir with mass-loss scaling with the SFR, links together three different aspects of the evolving galaxy population – (a) the cosmic time evolution of the specific star-formation rate sSFR relative to the growth of haloes, (b) the gas-phase metallicities across the galaxy population and over cosmic time, and (c) the ratio of the stellar to dark matter mass of haloes. The gas-regulator is defined by the gas consumption timescale ($\varepsilon^{-1}$) and the mass-loading $\lambda$ of the wind outflow $\lambda$·SFR.   The simplest regulator, in which $\varepsilon$ and $\lambda$ are constant, sets the sSFR to exactly the specific accretion rate of the galaxy: more realistic situations lead to an sSFR which is perturbed from this precise relation. Because the gas consumption timescale is shorter than the timescale on which the system evolves, the metallicity $Z$ is set primarily by the instantaneous operation of the regulator system rather than by the past history of the system. The metallicity of the gas reservoir depends on $\varepsilon$, $\lambda$ and the sSFR, and the regulator system therefore naturally produces a $Z(m_{star}, SFR)$ relation. Furthermore, this will be the same at all epochs unless the parameters $\varepsilon$ and $\lambda$ themselves change with time, naturally producing a so-called "fundamental metallicity relation".  The overall mass-metallicity relation $Z(m_{star})$ gives directly the fraction $f_{star}(m_{star})$ of incoming baryons that are being transformed into stars.  The observed $Z(m_{star})$ relation of SDSS galaxies implies a strong dependence of stellar mass on halo mass that reconciles the different faint end slopes of the stellar and halo mass-functions in standard $\Lambda$CDM. It also boosts the sSFR relative to the specific accretion rate and produces a different dependence on mass, both of which are observed.   The derived $Z(m_{star}, SFR)$ relation for the regulator system is fit to published $Z(m_{star}, SFR)$ data for the SDSS galaxy population, yielding $\varepsilon$ and $\lambda$ as functions of $m_{star}$. The fitted $\varepsilon$ is consistent with observed molecular gas-depletion timescales in galaxies (allowing for the extra atomic gas) while the fitted $\lambda$ is also plausible. The gas regulator model also successfully reproduces the $Z(m_{star})$ metallicities of star-forming galaxies at $z \sim 2$.  One consequence of this analysis is that it suggests that the $m_{star}$-$m_{halo}$ relation is established by baryonic processes operating *within* the galaxies, and that a significant (40%), and more or less constant, fraction of baryons coming into the halos are being processed through the galaxies. The success of the gas regulator model in simultaneously explaining so many diverse observed relations over the $0 < z < 2$ interval suggests that the evolution of galaxies is governed by the simple physics on which it is based.

# 1. Introduction

The goal of this paper is to explore and develop links between three different aspects of the evolving galaxy population.  These are:

(a)    The evolution with cosmic time of the rate at which stars are forming in galaxies, as characterized by the specific star-formation rate, sSFR, of Main Sequence galaxies and the relation of this to the growth of dark matter structures;

(b)    The dependence of the gas-phase chemical abundance of galaxies on their stellar mass, $m_{star}$, and star-formation rate, SFR, and the evolution of this $Z(m_{star}, SFR)$ relation with cosmic epoch;

(c)    The strong dependence of stellar mass on dark matter halo mass that is required to reconcile the faint end slopes of the observed galaxy mass function and the halo mass function expected from the standard $\Lambda$CDM cosmogony.

It will be shown that these three aspects of galaxy evolution are closely linked through the action of a single simple physical model for star-formation in galaxies in which the star-formation rate is instantaneously regulated by the mass of gas in the galaxy.

We stress at the outset that the goal is not to construct an accurate and therefore complex physical model of galaxies, to determine precise values of physical quantities from observations, or to try to rule out alternative models.  Rather, the approach is to take an extremely simple model and explore the consequences analytically so as to understand better how the different aspects of galaxy evolution, listed above, may be linked together in a single coherent view.  As a result, the construction of the model will involve a number of simple assumptions, any of which may be challenged in detail, but which will be adopted on a heuristic basis.

## 1.1    *The sSFR of star-forming galaxies and its evolution with time*

Out to $z \sim 2$, there is now good evidence for a "Main Sequence" of star-forming galaxies, in which the SFR is closely correlated to the existing stellar mass $m_{star}$ of the galaxy, with a scatter of only about 0.3 dex[1] around the mean relation. The Main Sequence has a characteristic sSFR that declines weakly with stellar mass as $sSFR \propto m_{star}^{\beta}$ with $\beta \sim -0.1$ (see e.g. Noeske et al 2007, Daddi et al 2007a, Elbaz et al 2007, Peng et al 2010). About 1%-2% of star-forming galaxies lie above the Main Sequence with significantly elevated star-formation rates. These may be the result of a major merger or other event but, over the $0 < z < 2$ range, these starbursts only contribute of order 10% of the total star-formation (Sanders & Mirabel 1984, Sanders et al 1988, Rodighiero et al 2011, Sargent et al 2012). There is also a substantial population of "quenched" passive galaxies whose sSFR are substantially lower than Main Sequence galaxies. These passive galaxies dominate the galaxy population at high stellar masses but the star-formation in them can be neglected for most purposes. Most stars therefore form in the Main Sequence galaxies that are the subject of this paper.

---

[1] Throughout this paper we will use dex to refer to the anti-logarithm, i.e. 0.3 dex corresponds to a factor of 2.

It is clear that the characteristic sSFR of the Main Sequence population evolves strongly with redshift, increasing by a factor of 20 back to $z \sim 2$, i.e. going roughly as $sSFR \propto t^{-2.2}$, where $t$ is the cosmic epoch (Elbaz et al 2007, Daddi et al 2007, Panella et al 2009), or as $sSFR \propto (1 + z)^3$. The behavior at higher redshifts is less clear: initial evidence that the sSFR levels off dramatically at $z > 2$ (Gonzalez et al 2010) has been challenged (Schaerer et al 2012, Stark et al 2012). The evolution above $z \sim 2$ does appear to flatten, and may be more like $(1+z)^{1.5}$. In fact, the existence of a tight Main Sequence is not well established at these earlier epochs.

A small complication in considering the sSFR is that some fraction $R$ of the mass that is converted into stars, as measured by the SFR, is promptly (we will assume instantaneously) returned to the interstellar medium, with the remaining $(1-R)$ staying in the form of long-lived stars. The build-up of the long-lived stellar population therefore has a characteristic timescale that is given by the inverse of a *reduced* specific star-formation rate (*rsSFR*), which is smaller than the *sSFR* by a factor $(1-R)$. Assuming an instantaneous mass return of $R$,

$$rsSFR = (1 - R) \cdot sSFR. \qquad (1)$$

It should be noted that there are different conventions in the literature for the calculation of stellar masses from spectrophotometric data combined with population synthesis models, and thus for the computation of sSFR. Often (e.g. as in Panella et al 2009) the stellar mass is the "actual" stellar mass of surviving long-lived stars (plus stellar remnants) derived from a stellar population model. An sSFR computed in this way requires a correction to yield the rsSFR. Alternatively, the stellar mass may be computed from the integral of the SFR of a particular stellar population model (e.g. as in Ilbert et al 2009). An sSFR computed from these values would already be equivalent to the rsSFR defined in Equation (1). We will adopt in this paper the convention that the sSFR is the SFR divided by the *actual* stellar mass, and will use the reduced rsSFR when needed to describe the e-fold time of the long-lived stellar population.

With the instantaneous return assumption, and taking $R = 0.4$ from stellar population models (e.g. Bruzual & Charlot 2003), and based on the data from Noeske (2007), Elbaz et al (2007), Daddi et al (2007), Panella et al (2009), Stark et al (2012) we adopt an $sSFR(m_{star},z)$ relation of the form (see Figure 1)

$$rsSFR(m_{star},t) = 0.07 \left(\frac{m_{star}}{10^{10.5}M_\odot}\right)^{-0.1} (1+z)^3 \text{ Gyr}^{-1} \qquad \text{(at } z < 2)$$

$$= 0.3 \ \left(\frac{m_{star}}{10^{10.5}M_\odot}\right)^{-0.1} (1+z)^{5/3} \text{ Gyr}^{-1} \qquad \text{(at } z > 2) \qquad (2)$$

The average specific accretion rate, or specific "mass increase rate", of dark matter haloes, which we denote sMIR$_{DM}$ (see Equation 17 below), is given by Neistein & Dekel (2008) as

$$sMIR_{DM} = 0.027 \left(\frac{m_{halo}}{10^{12}M_\odot}\right)^{0.15} (1+z+0.1(1+z)^{-1.25})^{2.5}$$

$$\sim 0.036 \left(\frac{m_{halo}}{10^{12}M_\odot}\right)^{0.15} (1+z)^{2.35} \qquad (3)$$

The similarities between the average observed rsSFR and the average sMIR$_{DM}$ on galactic mass scales from Equations (2) and (3) are shown in Figure 1, where we plot the cosmological evolution of both quantities with epoch for $m_{halo} = 10^{11.5}M_\odot$ and $m_{star} = 10^{10}M_\odot$ (left panel) and the variation with mass at two particular epochs, $z = 0$ and $z = 2$ (right panel). As has been noted before, the evolution of both quantities is very similar.

As well as broad similarities, there are however significant differences between these two quantities, as seen on Figure 1. The sSFR appears to be consistently higher than the sMIR$_{DM}$ over a wide range of redshifts, implying a shorter e-fold time for the build-up of stars compared with that of the dark matter haloes. Also, the slope of the (weak) mass-dependence of the sSFR and sMIR$_{DM}$ is reversed, with a logarithmic slope $\beta_{DM} \sim +0.15$ compared with the $\beta_{sSFR} \sim -0.1$. Given the overall similarities, the origin of these differences is something of a puzzle (e.g. Bouche et al 2010, Weinmann et al 2011).

## 1.2    Gas phase metallicities within the galaxy population

It has been known for many years that there is a significant range of metallicity within the galaxy population.  In this paper we are primarily concerned with the gas phase metallicities, $Z$, that are derived for star-forming galaxies from line ratios in HII regions. These reflect the metallicity of the gas out of which stars are being formed.

There are uncertainties in the estimation of gas-phase metallicities from emission line spectra (see e.g. discussion in Kewley & Dopita 2002, Kewley & Ellison 2008, Andrews & Martini 2012, Yates et al 2012) and different analyses produce significantly different metallicities, even from the same input data.  There is, however, no dispute that there is a strong overall trend with the stellar mass of the galaxy (Lequeux et al. 1979), well illustrated by the SDSS $Z(m_{star})$ mass-metallicity relation presented in Tremonti et al. (2004, hereafter T+04).  There is also good evidence for an evolution in the mass-metallicity relation with redshift (e.g. Savaglio et al. 2005, Maier et al., 2006, Erb et al. 2006, Maiolino et al. 2008) although the available emission line data at high redshifts is much more limited. Not least the extensive Erb et al data were based on the somewhat problematic NII/H$\alpha$ ratio.   Explanations of the existence of the mass-metallicity relation are still debated, but have included differing star-formation efficiencies within galaxies (e.g. Brooks et al 2007, Mouhcine et al 2008, Calura et al., 2009), supernova-driven winds (e.g. Larson 1974, T+04, Dalcanton, 2007, Finlator & Davé 2008), and variations in the initial mass function of stars (Köppen et al 2007).

Based on the extensive SDSS data, it has also been claimed that $Z$ correlates with other galactic parameters, most notably that there is an anti-correlation with the SFR of the galaxies (Ellison et al 2008, Mannucci et al 2010, hereafter M+10, Lara-Lopez et al 2010, Andrews & Martini 2012), especially at low galactic masses ($m_{star} \leq 10^{10}$ M$_\odot$).  In particular, M+10 presented a $Z(m_{star}, SFR)$ relation (reproduced below in Figure 5) and furthermore claimed that high redshift galaxies follow exactly the same relation, coining the phrase "fundamental metallicity relation" (FMR).   Other studies of the stability of the FMR with epoch have been undertaken by Richard et al (2012), Nakajima et al (2012), Cresci et al (2012), and Dessauges-Zavadsky et al (2012) amongst others.

While the existence and sign of the effect is quite well established at low masses (and low redshifts), the amplitude of the variation with SFR is quite uncertain, being reported between 0.2-0.6 dex, or even larger at low stellar masses (Andrews & Martini 2012). At high masses, the effect is much smaller and may vanish or even reverse (Yates et al 2012).   Explanations for why there might be a second SFR parameter in the mass-metallicity relation have hitherto generally involved *ad hoc* descriptions of adding pristine gas to boost star-formation (e.g. M+10, Dayal et al., 2012).

*1.3 The stellar mass vs. halo mass relation*

The theory of structure formation in the $\Lambda$CDM cosmology has proved very successful. A lingering issue has however concerned the faint end slope of the galaxy (stellar) mass function, as described by the Schechter parameter $\alpha$, when compared with the predicted slope of the dark matter halo mass function. These are significantly different, over several decades of stellar mass. This implies a strong dependence of the stellar mass on the halo mass (Vale & Ostriker, 2004), extending all the way up to the galaxy masses around M*, the characteristic knee in the Schechter function, at which point the relation turns over and $m_{star}/m_{halo}$ decreases. This latter effect is due to the quenching of star-formation in massive galaxies, which is not considered in this paper.

In the power-law regime at relatively low masses, the inferred halo mass dependence is sensitive to the measured faint-end slope $\alpha$ of the galaxy mass function (see Equation 8 below). Parameterizing $m_{star} \propto m_{halo}{}^{\gamma}$, values of $\gamma$ as high as $\gamma \sim 3$ have been inferred (Guo et al, 2010). Using the Schechter faint-end slope $\alpha$ for star-forming galaxies which is $\alpha = -1.45 \pm 0.05$ (Peng et al 2010, Baldry et al 2012) yields $m_{star} \propto m_{halo}{}^{1.85 \pm 0.2}$ for $\alpha_{DM} \sim -1.83$. Although less extreme, this still implies a rather dramatic breaking of the coupling between halo mass and stellar mass that sits uncomfortably with the impression of the close coupling between the two that comes from the comparison of the galactic sSFR and the halo sMIR$_{DM}$ that was outlined in Section 1.1.

The strong dependence of stellar mass on halo mass at low masses has been thought to reflect either the action of winds leading to baryonic mass loss (Larson 1974, White & Rees 1978) or a variation in the efficiency with which galaxies convert gas into stars. The link with metallicity has previously been explored by several authors (e.g. Finlator & Davé 2008, Peeples & Shankar 2010).

*1.4 The gas-regulated model of galaxies*

The physical model of this paper is built around the close coupling of baryons to dark matter that is indicated by the similarity of Equations (2) and (3) illustrated in Figure 1. The goal is to develop a very simple physical model that is motivated by this observation plus the application of reasonable physical assumptions, and see how far we can go in linking together the different issues that we have outlined in the previous three sections. In developing this analysis, we will make a number of heuristically-motivated simplifying assumptions whose individual detailed validity can obviously be questioned. We again stress that the goal is not to construct a physically detailed model that can be quantitatively compared to accurate data to determine physical parameters, but rather to provide a framework for considering the general problems in galaxy evolution.

The analysis will be built around a model for individual galaxies in which a given galaxy is continuously fed from outside by new gas, as in the model of Bouché et al (2010). The SFR in the galaxy is set by the instantaneous mass of gas within the galaxy, i.e. the gas content of a galaxy regulates its SFR. If the mass-loss in winds in a given system is proportional to the SFR, then the regulator keeps the SFR as a constant, or slowly evolving, fraction $f_{star}$ of the baryons flowing into the galaxy "system". The "system" considered in this Paper will consist of the stellar component of the galaxy plus the gas in the reservoir within the galaxy. The gas within this internal reservoir takes part in the chemical mixing of the galaxy and it is this reservoir of gas that sets and regulates the SFR of the galaxy. The system is however open: gas may flow in from the surroundings (i.e. from further out in the halo) and out again in a star-formation driven wind. However, it is assumed that the reservoir is isolated from the surrounding gas in the halo except via these well-defined flows in and out of the system. We will also not consider chemical mixing between the outflowing and inflowing material.

This very simple model for a galaxy has the interesting effect of setting the rsSFR to be close to the specific infall rate of gas into the system. The regulator is an entirely baryonic system. The *baryonic* specific infall rate, which we designate as the sMIR$_B$, is the rate of inflow of gas divided by the integral of that rate over all previous time (see Equation 16 later in the paper). The integral may not be the actual baryonic mass of the system, because some of the previously accreted baryonic material may have been lost through winds. The sMIR$_B$ of a galaxy is not therefore an observable quantity, but it would be expected to be closely linked to the sMIR$_{DM}$ of its parent halo. In fact, the two will be identical if a constant fraction of the baryons entering the halo penetrate down to enter the galaxy system considered above (assuming a cosmic baryon to dark matter fraction enters the halo).

The smooth sMIR$_{DM}$ given by Equation (3) represents an average across the population. In practice, the accretion of dark matter and baryons onto a given halo may proceed in a more irregular way as lumps of material are brought in. However, the subsequent accretion of gas onto the central galaxy will be smoothed out, at least on the halo dynamical timescale of order 0.1 $t_H$, and possibly longer. A key part of our model is that, at any epoch, the spread in the sSFR on the Main Sequence reflects the range of smoothly varying accretion rates of gas onto galaxies.

*1.5 Layout of the Paper*

The layout of the paper is as follows: In Section 2 we develop the basic features of the simple gas-regulated model, emphasizing its action in terms of the resulting sSFR and the ratio μ of gas to stars in the system. We also look at the different timescales in the system and the ability of the regulator to smoothly control the SFR in a galaxy. We then derive in Section 3 the key expression(s) for the instantaneous gas metallicity of the gas in the reservoir and also consider how this may change with time. We will find that the action of the gas regulator naturally produces a $Z(m_{star}, SFR)$ relation that will be the same at all epochs unless the parameters describing the internal action of the regulator change. We also show that the metallicity of the system is very simply related to the fraction, $f_{star}$, of baryons that enter the system and are transformed into long-lived stars. In Section 4 we show that the mass-metallicity relation of SDSS galaxies implies an $f_{star}(m_{star})$ that increases with mass and that this naturally accounts for the different faint end slopes of the galaxy and halo mass functions. Also, since a given galaxy will increase its $f_{star}$ as it grows in mass, this will produce an rsSFR that will be systematically higher than the $sMIR_B$, and thus likely the $sMIR_{DM}$. It will also produce a different mass-dependence of rsSFR$(m_{star})$ compared with $sMIR_{DM}(m_{halo})$ Finally, in Section 5, we compare the $Z(m_{star}, SFR)$ metallicity relation from our regulator model with the $Z(m_{star}, SFR)$ data for the SDSS population that has been presented by M+10, and the Tremonti et al (2004, T+04) $Z(m_{star})$ relation. By fitting these data, we derive the mass-dependences of star-formation efficiency and wind mass-loading that are required to reproduce the metallicity data across the galaxy population. These quantities are found to be broadly reasonable. Taking them at face value, we then look at the division of incoming baryons between stars, the reservoir and the outflow at $z = 2$ and $z = 0$, and make a prediction for the form of the $Z(m_{star})$ relation at $z = 2$ which is largely consistent with observations. Section 6 of the paper then summarizes the key points from the earlier sections.

Finally, it should be noted that stellar masses and star-formation rates taken from the literature and are for a concordance cosmology with $H_0 \sim 70$ kms$^{-1}$Mpc$^{-1}$. We will assume when necessary that baryons and dark matter are well-mixed flowing into haloes, with a cosmic fraction of baryons $\Omega_B/\Omega_M = 0.15$.

## 2. The ideal gas-regulated model

### 2.1 The ideal operation of the gas-regulator

The most important features of the gas-regulated model of galaxies that we develop in this paper are very simple. The model assumes that gas, mixed with dark matter, flows from the surroundings into a halo. Some fraction $f_{gal}$ of the incoming baryons penetrate down to enter the galaxy system as baryonic gas. There the gas adds to a reservoir within the galaxy. The instantaneous SFR in the galaxy is determined by the instantaneous mass of gas in this internal reservoir. Metals are returned to this internal reservoir. Finally, some gas may be expelled from the reservoir back out to the halo, or even beyond, in a galactic wind.

We first define the star-formation efficiency ε of a particular galaxy system in terms of the instantaneous SFR and the mass of gas present in the reservoir within the system, $m_{gas}$, as

$$SFR = \varepsilon m_{gas}. \tag{4}$$

The gas consumption timescale $\tau_{gas}$ is then simply given by the inverse of ε, i.e.

$$\tau_{gas} = \varepsilon^{-1}. \tag{5}$$

It should be noted that in this analysis, $m_{gas}$ is the total mass of gas within the gas reservoir of the galaxy and not just the molecular gas out of which stars are actually being formed. Correspondingly, $\tau_{gas}$ will be longer than the $\tau_{gas}$ derived from the molecular gas alone.

In a very simple regulator model, the star-formation efficiency ε would be a constant. However, for realistic galaxies, ε will likely be a function of the overall mass of the galaxy, or possibly of other galactic parameters such as the dynamical timescale. In what follows, ε will be taken to be a general $\varepsilon(m_{star}, t)$ function, but we will assume that it varies on a timescale that is longer than $\tau_{gas}$.

Defining the mass ratio of gas to stars in the system to be μ,

$$\mu = \frac{m_{gas}}{m_{star}}, \tag{6}$$

then Equation (4) can be re-written in terms of the sSFR and ε:

$$\mu = \varepsilon^{-1} \cdot sSFR \quad . \tag{7}$$

Note that in a more general case where the SFR varies as some power κ of the gas mass, SFR = ε·$m_{gas}^{\kappa}$, as in a Kennicutt-Schmidt-type relation (Kennicutt 1998), then Equation (7) takes the form

$$\mu = (\varepsilon^{-1} \cdot sSFR)^{1/\kappa} \cdot m_{star}^{-\left(\frac{\kappa-1}{\kappa}\right)}. \tag{7b}$$

It should be noted that the regulation of the star-formation rate by the gas reservoir will occur for any positive value of κ, i.e. we simply need the SFR to increase if the mass of gas present increases. Adopting κ = 1 is a simple case that we adopt for heuristic purposes.

We define the mass-loss Ψ from the system, which we assume increases (in a given system) linearly with the SFR, in terms of a mass-loading factor λ, so that

$$\Psi = \lambda \cdot SFR \tag{8}$$

Again, λ may well vary with the mass of the galaxy (and/or epoch) due to the depth of the potential well, or other factors. Note that the gas-to-stars ratio μ in the steady-state regulator system depends only on the sSFR and on the star-formation efficiency ε, through Equation (7), and not on the level of mass-loss from the system. Again, the linearity of Ψ with respect to the SFR in a particular gas is a simplifying, heuristic, assumption. We will see that the outflow acts as a kind of inefficiency in the system, and its exact form will not alter the qualitative features of the model.

The simplest possible case would be one in which the ε and λ of a given regulator system are both constant. We will refer to this as an "ideal" regulator. In practice we expect that ε and λ will both depend on galactic mass, and possibly on epoch, and these parameters will change for a given galaxy as it increases in mass, even if at fixed mass these parameters are constant with epoch. However, provided changes to the operation of the regulator are slow compared with the gas consumption timescale $\tau_{gas}$, then the ideal case will be a good basis for considering these more realistic situations the outcome is perturbed (see Sections 4 and 5).

The action of the basic Equation (4) is to regulate the SFR in a galaxy via the amount of gas present. Variations in the infall rate, over time or from galaxy to galaxy (or, equally well, variations in the star-formation efficiency or wind mass-loading) will quickly lead to changes in the gas reservoir, consequent adjustment of the SFR, and thus *regulation* of the star-formation rate.

Two simple diagrammatic representations of the gas-regulated model are shown in Figure 2. The one on the left is more pictorially realistic, while the one on the right shows a more schematic representation of the flows through the system. Gas flows into this reservoir from outside, at a rate given by $\Phi$. In a given interval of time, some of the gas in the reservoir is formed into stars, and a fraction $(1-R)$ of this steadily builds up a population of long-lived stars. Star-formation may drive a wind $\Psi$ out of the galaxy, either back into the halo, or beyond (we will not be concerned which). The mass of gas in the reservoir of the system, $m_{gas}$, is free to increase or decrease with time and it is this change which gives the regulator its ability to regulate the SFR of the galaxy. Changes in $m_{gas}$ must be associated with a net flow into or out of the reservoir. We will not consider the gas in the wider halo except to define the instantaneous inflow $\Phi$ of gas into the galaxy to be some fraction $f_{gal}$ of the instantaneous inflow of baryons into the halo. The gas flowing into the galaxy system from the surrounding halo may have some prior chemical abundance $Z_0$, and the gas flowing back out is assumed to have the composition representative of the reservoir. No attempt will be made to follow possible mixing of these two flows in the surrounding halo environment.

Strict mass conservation, plus our definition of $\Phi$ in terms of the increase in halo mass and $f_{gal}$, enables us to write

$$\Phi = (1 - R + \lambda) \cdot SFR + \frac{dm_{gas}}{dt} \quad . \tag{9}$$

In Davé et al's (2012) recent treatment, the $dm_{gas}/dt$ term in Equation (9) is set to zero, since they assumed that the gas reservoir has a fixed mass. Changes in $m_{gas}$ are a fundamental part of the self-regulation of the galaxy in our model. Our gas-regulator model will have a stable (or slowly evolving) gas *ratio* $\mu$ if the sSFR is more or less constant. We will see below that the increase of the gas reservoir can actually be the *dominant* destination of incoming baryons at $z \sim 2$ in galaxies of intermediate masses. More importantly, we shall also see that it is the inclusion of this variable reservoir term that leads to the implicit dependence of gas metallicity on the SFR, which is otherwise not present.

The change in gas mass can be expressed in terms of the SFR and the change in μ as follows:

$$\frac{d\mu}{dt} = -\frac{m_{gas}}{m_{star}^2}(1-R)SFR + \frac{1}{m_{star}}\frac{dm_{gas}}{dt},$$

so

$$\frac{dm_{gas}}{dt} = \mu(1-R)SFR + m_{star}\frac{d\mu}{dt}.$$

Since $m_{star} = \mu^{-1}m_{gas}$ and $m_{gas} = \varepsilon^{-1} \cdot$SFR, this becomes

$$\frac{dm_{gas}}{dt} = \left(\mu(1-R) + \varepsilon^{-1}\frac{d\ln\mu}{dt}\right) \cdot SFR. \tag{10}$$

The last term in the bracket in Equation (10) is the ratio between the gas consumption timescale $\tau_{gas}$ and the timescale on which μ, i.e. the ($\varepsilon^{-1} \cdot$sSFR) product from Equation (7), is changing. A condition for the regulator to achieve a quasi-steady state is that this ratio should be small, i.e. that $\tau_{gas}$ should be shorter than the timescale on which μ ($\tau_{gas} \cdot$sSFR) is changing. In the simplest "ideal" regulator, in which ε and the input $sMIR_B$ are both constant, this term will of course be identically zero. We will return to consider this quantity in more realistic situations in Section 5.

Substituting Equation (10) into Equation (9) then yields an expression linking the inflow rate Φ and the SFR

$$\Phi = \left((1-R)(1+\mu) + \lambda + \varepsilon^{-1}\frac{d\ln\mu}{dt}\right) \cdot SFR. \tag{11}$$

With our heuristic assumptions of $SFR \propto m_{gas}$ and $\Psi \propto SFR$, all of the flows in the system scale with the SFR (and Φ) and a constant fraction $f_{star}$ of baryons are being transformed into stars. The fractional splitting of the incoming baryons is given by the relative values of ε, λ and the sSFR and may be represented by $f_{star}$, $f_{out}$ and $f_{res}$ respectively (see Figure 2), which are given as exact solutions from Equation (11) as

$$f_{star} = \frac{(1-R) \cdot SFR}{\Phi} = \frac{1}{1+(1-R)^{-1}\lambda+\mu+(1-R)^{-1}\varepsilon^{-1}\frac{d\ln\mu}{dt}} = \frac{1}{1+(1-R)^{-1}\lambda+\varepsilon^{-1}\left(sSFR+(1-R)^{-1}\frac{d\ln\mu}{dt}\right)} \tag{12}$$

$$f_{out} = \frac{\lambda \cdot SFR}{\Phi} = \frac{(1-R)^{-1}\lambda}{1+(1-R)^{-1}\lambda+\mu+(1-R)^{-1}\varepsilon^{-1}\frac{dln\mu}{dt}} = \frac{(1-R)^{-1}\lambda}{1+(1-R)^{-1}\lambda+\varepsilon^{-1}\left(sSFR+(1-R)^{-1}\frac{dln\mu}{dt}\right)} \qquad (13)$$

$$f_{res} = \frac{\left(\mu\,(1-R)+\varepsilon^{-1}\frac{dln\mu}{dt}\right)\cdot SFR}{\Phi} = \frac{\mu+(1-R)^{-1}\varepsilon^{-1}\frac{dln\mu}{dt}}{1+(1-R)^{-1}\lambda+\mu+(1-R)^{-1}\varepsilon^{-1}\frac{dln\mu}{dt}} = \frac{\varepsilon^{-1}sSFR+(1-R)^{-1}\varepsilon^{-1}\frac{dln\mu}{dt}}{1+(1-R)^{-1}\lambda+\varepsilon^{-1}\left(sSFR+(1-R)^{-1}\frac{dln\mu}{dt}\right)}$$

$$(14)$$

which simplify to the following if we neglect the slow time dependence of μ, as in the case of the "ideal" regulator (but see Section 5 below for more realtistic situations):

$$f_{star} = \frac{(1-R)\cdot SFR}{\Phi} = \frac{1}{1+(1-R)^{-1}\lambda+\mu} = \frac{1}{1+(1-R)^{-1}\lambda+\varepsilon^{-1}\cdot sSFR} \quad, \qquad (12a)$$

$$f_{out} = \frac{\lambda\cdot SFR}{\Phi} = \frac{(1-R)^{-1}\lambda}{1+(1-R)^{-1}\lambda+\mu} = \frac{(1-R)^{-1}\lambda}{1+(1-R)^{-1}\lambda+\varepsilon^{-1}\cdot sSFR} \quad, \qquad (13a)$$

$$f_{res} = \frac{\mu\,(1-R)\cdot SFR}{\Phi} = \frac{\mu}{1+(1-R)^{-1}\lambda+\mu} = \frac{\varepsilon^{-1}sSFR}{1+(1-R)^{-1}\lambda+\varepsilon^{-1}\cdot sSFR} \quad. \qquad (14a)$$

Clearly, $f_{star} + f_{out} + f_{res} = 1$ in both sets of equations, as required from mass conservation. It should be noted that the reservoir can be a significant or even dominant destination of baryons, especially if the gas ratio μ is high as it will be at high redshift because of the high sSFR. Finally, it is trivial to compute the gas fraction of the system $\nu_{gas}$ directly from the gas ratio μ given in Equation (7),

$$\nu_{gas} = \frac{m_{gas}}{m_{gas}+m_{star}} = \frac{\mu}{1+\mu} = \frac{1}{1+\varepsilon\cdot sSFR^{-1}} \quad. \qquad (15)$$

The net effect of these flows of gas through the regulator system is shown schematically on the right-hand side of Figure 2. Provided the gas consumption timescale $\tau_{gas}$ in the galaxy is short compared with the timescale on which the outside conditions and the parameters of the regulator are varying (as will be discussed further below), gas can be viewed as continuously flowing through the system in a quasi steady-state. The flow of incoming gas Φ divides into three: a flow into long lived stars, one (which can flow in either direction) that changes the mass of gas within the reservoir, and one that takes gas out of the system. The three-way dividing action of the regulator system is set by just three parameters, the star-formation efficiency ε, the mass-loading factor λ and the sSFR (which is set by the input sMIR$_B$) which gives the required μ via ε as in Equation (7).

## 2.2 Action of the ideal regulator in terms of the specific inflow rate

The ideal regulator is a system with constant ε and λ, and in which the input specific accretion rate varies on a timescale that is long compared with the gas consumption timescale. As introduced earlier, the specific inflow rate of the gas, sMIR$_B$, is defined to be the ratio of the instantaneous gaseous inflow rate Φ to the past integral of Φ over all time, i.e.

$$sMIR_B = \frac{\Phi(t)}{\int_0^t \Phi(t')dt'}.$$  (16)

It should be noted that the denominator may not be the existing baryonic mass of the galaxy if mass has been lost from the system, and the sMIR$_B$ is therefore not a directly observable quantity. However, if we assume that the ratio of gaseous baryons to dark matter entering a halo is fixed (e.g. at the cosmic ratio), and if the fraction $f_{gal}$ of baryons that enter the halo and penetrate all the way down to enter the galaxy regulator system is also constant, then the sMIR$_B$ will be exactly equal to the specific mass increase rate of the dark matter halo, defined as

$$sMIR_{DM} = \frac{1}{m_{halo}} \frac{dm_{halo}}{dt}.$$  (17)

The sMIR$_{DM}$ is observable at least in the sense of being measurable in numerical simulations. The sMIR$_{DM}$ from simulations was plotted in Figure 1.

As noted above, the four flows that are shown in the right hand part of Figure 2 will all scale linearly with the SFR, and thus with the inflow Φ. The three branches, into which the flow of incoming baryons splits, depend on the values of ε, λ and the sSFR. In the ideal case, all of these will be constant. In a more realistic situation, they may change on a timescale that is long compared with the time τ$_{gas}$ that a given packet of gas stays in the system.

If ε, λ and the rsSFR are indeed constant, then it follows from Equation (12) that $f_{star}$ will be constant as well. Thus the ideal regulator transforms a constant fraction of the inflowing baryons into stars. As long as this occurs, then the sSFR will quickly converge to the sMIR$_B$ as soon as the build-up of stars has proceeded far enough that the initial conditions are forgotten. This can be seen as follows:

$$SFR = f_{star}\Phi$$

$$m_{star}(t) = m_{star}(t_0) + f_{star}\int_{t_0}^{t}\Phi\,dt$$

Once the initial mass of stars $m_{star}(t_0)$ is negligible compared with the new ones produced with the constant $f_{star}$, then dividing the first equation by the second gives

$$sSFR \approx sMIR_B \tag{18}$$

This exact convergence of the sSFR and the $sMIR_B$ requires that the parameters describing the action of the regulator, specifically the star-formation efficiency ε and the mass-loading λ, are constant (our "ideal" case). If these parameters change with time, e.g. either directly or via a mass-dependence as the galaxy grows, then the division of the incoming baryons, and thus $f_{star}$, will no longer be exactly constant. This will perturb the ideal equality between the sSFR and $sMIR_B$. We explore this in Sections 4 and 5.

Why do we focus on the sSFR and specific accretion rates rather than the straight-forward SFR and inflow rate Φ? As described above, the ideal regulator works by setting the SFR to some (constant) fraction $f_{star}$ of Φ, but Φ is, at least in practical terms, not an observable quantity. In contrast, the sSFR and $sMIR_{DM}$ are both readily observable (the latter in numerical simulations). Furthermore, while mass-loss λ and the star-formation efficiency ε will change $f_{star}$ (from Equation 12), it will have no effect on the equality of the sSFR and $sMIR_B$ produced by the regulator. We noted in Section 1.1 the strong empirical similarities between the observed sSFR of Main Sequence galaxies and the specific mass increase rate of dark matter haloes in numerical dark matter simulations. The action of the ideal regulator in forcing the sSFR to the $sMIR_B$ (and to the $sMIR_{DM}$ if $f_{gal}$ is constant) is therefore a strong argument in favor of exploring this simple gas-regulator system as a good working model for galaxy evolution.

### 2.3 Conditions for the regulator to operate

The operation of the ideal regulator defined by Equation (4) is shown in Figure 3 using a simple numerical model under different conditions. The parameter ε is taken to be constant. Gas is fed into the system at a rate that is given by the $sMIR_B$. Stars are then formed according to Equation (4) with a certain ε (or, equivalently, $\tau_{gas}$). The response of the ideal regulator system is shown in terms of the resulting sSFR (upper two panels) and the gas ratio μ (lower two panels). We here neglect mass-loss from the system since it will not affect either of these two quantities, as argued above, and we set $R = 0$ for simplicity. Four different scenarios have been chosen to illustrate features of the ideal regulator, and are not intended to be relevant for actual galaxies.

In the two left hand panels of Figure 3, we show a situation in which the input $sMIR_B$ is held constant at 1 Gyr$^{-1}$, except for an abrupt period during which it is raised by a factor of ten. In each case, we have plotted the outcomes with five different values of $\tau_{gas}$ (= ε$^{-1}$), spaced logarithmically from 0.1 - 10 Gyr (the longest $\tau_{gas}$ is indicated by the dashed lines, and the shortest by dotted ones) so as to span the case where $\tau_{gas} \gg sMIR_B^{-1}$ to $\tau_{gas} \ll sMIR_B^{-1}$. The figure shows how the system responds to the sharp changes in $sMIR_B$ and quickly adjusts the sSFR to its new value. Interestingly, the convergence timescale is set by the *shorter* of $\tau_{gas}$ and the sSFR$^{-1}$ timescale. It can be seen that the system has no trouble maintaining sSFR = $sMIR_B$ as long as the $sMIR_B$ is constant, even for the case where the $\tau_{gas}$ is very much longer than the *sMIR_B$^{-1}$* and sSFR$^{-1}$ timescales (e.g. the scenario shown by the dotted line). The bottom left panel shows the gas ratio μ that is produced for these same two scenarios. Clearly the equilibrium value of μ does depend, strongly, on ε and the sSFR, as expected from Equation (7).

Two further scenarios are shown in the two right hand panels to explore the effect of gradually changing the $sMIR_B$. The $sMIR_B$ is initially set to 1 Gyr$^{-1}$ (as in the left hand panels, so the curves are continuous across the divide), but it then accelerates away from this value, either to higher or lower values. In the former case, the ideal regulator is able to maintain sSFR = $sMIR_B$ for all $\tau_{gas}$. If however the $sMIR_B$ drops, as in the lower set of curves, then the regulator can only maintain sSFR = $sMIR_B$ for as long as $\tau_{gas}$ is shorter than the timescale on which the $sMIR_B$ is changing. If the $sMIR_B$ decreases too quickly, the system cannot respond fast enough and the sSFR breaks away from the $sMIR_B$ and declines with the $\tau_{gas}$ timescale. The asymmetry in behavior between an increasing and a decreasing $sMIR_B$ arises because, in this simple model, the SFR can increase instantaneously (in principle) because the gas content of the galaxy can increase instantaneously, but can only decline on the timescale that the gas content declines, which is set by the gas consumption timescale $\tau_{gas}$ (provided there are no other mechanisms for removing gas).

The timescale condition for the ideal regulator may be written in terms of the timescale $\tau_{sMIR_B}$ on which the input $sMIR_B$ is changing

$$\tau_{gas} < \left[\frac{1}{sMIR_B}\frac{d(sMIR_B)}{dt}\right]^{-1} = \tau_{sMIR_B} \tag{19}$$

If, the sMIR is rising, then the sMIR (or sSFR) may be substituted for the $\tau_{gas}$ in Equation (19).

However, we commented above that in more realistic (i.e. "non-ideal") systems, the internal parameters of the regulator $\varepsilon$ and $\lambda$ may well depend on stellar mass and therefore with time as a galaxy grown in mass. Furthermore, the SFR efficiency $\varepsilon$ may change with cosmic epoch, even at fixed stellar mass, since $\tau_{gas}$ (at least for molecular gas) is observed to be shorter at high redshifts (Daddi et al 2010, Genzel et al 2010, Tacconi et al 2012). This may reflect a change in the dynamical times within galaxies. If $\varepsilon$ and $\lambda$ are indeed changing within the regulator system, then we will have an additional timescale constraint: the timescale on which $\varepsilon$ and $\lambda$ are changing must also be longer than the gas consumption timescale. Since the former is likely to be the timescale on which the stellar mass is increasing, i.e. $rsSFR^{-1}$, the regulator system may be unable to follow these changes if $\tau_{gas} > rsSFR^{-1}$.

### 2.4 Timescales for galaxy evolution

Since the star formation efficiency $\varepsilon$ (or $\tau_{gas}^{-1}$) presumably reflects the physics of star-formation and is not directly related to the growth of structure in the Universe, the condition Equation (19) is not trivially satisfied. We show in Figure 4, a number of timescales that will be relevant for typical massive galaxies ($m_{star} \sim 10^{10} M_{\odot}$) over cosmic time. Specifically, we plot the halo growth timescale $sMIR_{DM}^{-1}$ for $10^{11.5} M_{\odot}$ haloes from cosmological simulations (Equation 3), the stellar mass growth timescale $rsSFR^{-1}$ at $10^{10} M_{\odot}$ from observations (Equation 2), the observed molecular gas depletion timescale $\tau_{gas}$ (from Daddi et al 2007b, Genzel et al 2008 and references therein), the Hubble timescale $\tau_H$, i.e. the local age of the Universe, and finally the timescales on which the $sMIR_{DM}$ and $rsSFR$ are themselves changing, following the definition in Equation (19), which we denote as $\tau_{sMIR}$ and $\tau_{rsSFR}$ respectively. We also show the dynamical timescale $\tau_{dyn}$ of dark matter haloes, which is $\tau_{dyn} \sim 0.1 \, \tau_H$. The dynamical timescales within galaxies will be an order of magnitude smaller. Other timescales may also be relevant, including the time taken to produce the metals within the stellar population, to spread and mix the returned metals across the galaxy, for the gas to cool and form stars and for the new metals will be observable in the emission lines. A detailed treatment of this cycle is beyond the scope of this heuristic exploration of the gas-regulated model which is rather concerned with the quasi steady-state of a system in which we assume instantaneous recycling of the material.

It can be seen that $\tau_{gas}$ is comfortably shorter than the $\tau_{sMIR}$ for all epochs, and so the condition given by Equation (19) should be satisfied over essentially all of cosmic time if the gas supply follows the global evolution of the $sMIR_{DM}$ (we here do not consider cataclysmic events like mergers). However, we note that if $\tau_{gas}$ increases to lower masses then low mass galaxies may no longer satisfy this condition. In particular, if $\tau_{gas}$ scales as $m_{star}^{0.3}$ (see Section 5) then dwarf galaxies with $m_{star} \sim 10^8 \, M_\odot$ may not be effectively regulated.

While the timescale on which the *external* feeding of the regulator changes (for massive galaxies and ignoring mergers) should always comfortably longer than $\tau_{gas}$, Figure 4 also shows that the timescale on which the *internal* parameters of the regulator may be changing (if they are mass-dependent), i.e. the inverse of the specific star-formation rate, rsSFR$^{-1}$, may be comparable to the $\tau_{gas}$ at $z > 2$ (we here neglect non-molecular gas). Furthermore, at this point the dynamical time $\tau_{dyn}$ of the halo will also be comparable. It is not clear what the consequences of this convergence of timescales will be: if $f_{star}$ is increasing with stellar mass, as is likely (see below), then the SFR should be able to instantaneously adjust upwards, as discussed above. Situations in which $f_{star}$ is quickly decreasing would be more problematic as the SFR cannot adjust downwards faster than the $\tau_{gas}$ timescale. Nonetheless, the empirical convergence of the $\varepsilon^{-1}$ (i.e. $\tau_{gas}$), rsSFR$^{-1}$ and $\tau_{dyn}$ timescales at $z \sim 2$ may represent a natural transition point in the evolution of massive galaxies. It may well explain the possible change of behavior at $z \sim 2$ in the rsSFR($z$) evolution that was discussed in the Introduction, and which is also apparent on Figure 4, and may also be linked to the evidence of large-scale disk instabilities at $z > 2$ (Genzel et al 2008). In this regard, establishing the existence of a Main Sequence and determining its characteristic rsSFR, and empirically determining the $\tau_{gas}$ at $z > 2$ and establishing the mass-metallicity relation at these redshifts, will all be of great interest since these are all signatures of the gas-regulation of galaxies.

## 3 Metallicity in the gas-regulated model

We now determine the metallicity of the gas reservoir that sustains the regulator. This is a standard analysis and the derived relation is a special case of more general derivations (Recchi et al., 2008, Spitoni et al. 2010, Dayal et al. 2012). It is here reproduced for the particularly simple and distinctive regulator model that we are considering, as this imposes important linkages between infall, star-formation, mass-loss and the level of the gas-reservoir.

### 3.1 Metallicity within the regulator system

If the infalling gas has metallicity $Z_0$, then the change in the mass $m_Z$ of metals in the gas reservoir will be given in terms of the yield. We define the yield $y$ as the mass of metals returned to the interstellar medium per unit mass that is locked up into long-lived stars, i.e. (1-$R$) times the mass of stars formed. We can then write

$$\frac{dm_Z}{dt} = y(1-R) \cdot SFR - Z(1-R+\lambda) \cdot SFR + \Phi Z_0. \tag{20}$$

Eliminating $\Phi$ using Equation (11) we get

$$\frac{dm_Z}{dt} = \left( y(1-R) - (Z-Z_0)(1-R+\lambda) \right) \cdot SFR + Z_0 \frac{dm_{gas}}{dt}. \tag{21}$$

The change in metallicity $Z$ of the gas in the system is then just

$$\frac{dZ}{dt} = \frac{1}{m_{gas}} \left[ \frac{dm_Z}{dt} - Z \frac{dm_{gas}}{dt} \right],$$

so that

$$\frac{dZ}{dt} = \frac{1}{m_{gas}} \left[ \left( y(1-R) - (Z-Z_0)(1-R+\lambda) \right) \cdot SFR - (Z-Z_0) \frac{dm_{gas}}{dt} \right]. \tag{22}$$

or

$$\frac{dZ}{dt} = \left( y(1-R) - (Z-Z_0)(1-R+\lambda) \right)\varepsilon - (Z-Z_0)\frac{1}{m_{gas}}\frac{dm_{gas}}{dt}. \tag{23}$$

Inserting Equation (10) into Equation (23) and rearranging, we get

$$\varepsilon^{-1} \frac{dZ}{dt} = y(1-R) - (Z-Z_0)\left( 1-R+\lambda+\varepsilon^{-1}\cdot rsSFR + \varepsilon^{-1}\frac{d\ln\mu}{dt} \right). \tag{24}$$

As before, the final term will be small if the timescale for changes to μ (i.e. to the $\varepsilon^{-1}\cdot$sSFR product, from Equation 7) is longer than the $\tau_{gas} = \varepsilon^{-1}$ gas consumption timescale.

The action of equation (24) is therefore to rapidly drive the metallicity of the gas to an equilibrium value, $Z_{eq}$, that is given by setting $dZ/dt$ in Equation (24) to zero, i.e.

$$Z_{eq} = Z_0 + \frac{y(1-R)}{(1-R)+\lambda+\varepsilon^{-1}\cdot\left(rsSFR+\frac{d\ln\mu}{dt}\right)} \quad . \tag{25}$$

or

$$Z_{eq} = Z_0 + \frac{y}{1+\lambda(1-R)^{-1}+\varepsilon^{-1}\cdot\left(sSFR+(1-R)^{-1}\frac{d\ln\mu}{dt}\right)} \quad . \tag{26}$$

As shown in the Appendix, the timescale for driving $Z$ towards $Z_{eq}$ is of order $\tau_{gas}$, which is shorter than the timescale on which the equilibrium conditions, set by the sMIR$_B$, are varying. This is the justification for considering the gas metallicity to be instantaneously set by the parameters of the system as gas flows through it. The only knowledge of the history of the system is in the $d\ln\mu/dt$ term, which reflects changes to the ($\varepsilon^{-1}\cdot sSFR$) product (from Equation 7). As noted above, for the simplest "ideal" regulator this term will be zero. We will consider the value of this term in more realistic situations in Section 5. Here we note that, if the inflow is stopped, for example as part of a quenching process, then the SFR and μ will both decline exponentially towards zero with a timescale of (1-$R$)·$\tau_{gas}$ (see e.g. Figure 4). The first term in the bracket in Equation (26) will become negligible and the second term will become −1. The metallicity will rapidly increase, as in a closed box model, while preserving the (anti-) correlation with the SFR. We will not consider such quenching situations further in this Paper.

Using Equation (7), Equation (26) can also be re-written in terms of the gas fraction μ, as

$$Z_{eq} = Z_0 + \frac{y}{1+\lambda(1-R)^{-1}+\mu+(1-R)^{-1}\varepsilon^{-1}\frac{d\ln\mu}{dt}} \tag{27}$$

to recover a form that is similar to Equations (10) and (27) of Peeples & Shankar (2011). Using the definition of sSFR in terms of the SFR, this can be re-written explicitly in terms of the SFR

$$Z_{eq} = Z_0 + \frac{y}{1+\lambda(1-R)^{-1}+\varepsilon^{-1}\left(m_{star}^{-1}\cdot SFR+(1-R)^{-1}\frac{d\ln\mu}{dt}\right)} \quad . \tag{28}$$

Finally, $Z_{eq}$ can be written in terms of the ratio of $f_{star}$, which is the ratio of the reduced SFR to the infall rate $\Phi$:

$$Z_{eq} = Z_0 + y \frac{(1-R)SFR}{\Phi} = Z_0 + f_{star}\, y. \tag{29}$$

Not surprisingly, the metallicity $Z_0$ of the incoming gas just acts as an additive term in Equations (26-29). Note that, if the change in gas fraction $\mu$ can be neglected, as in the "ideal" regulator, then the $d\ln\mu/dt$ term in Equations (26-28) may be set to zero yielding

$$Z_{eq} = Z_0 + \frac{y}{1+\lambda(1-R)^{-1}+\varepsilon^{-1}\cdot sSFR} \quad , \tag{26a}$$

$$Z_{eq} = Z_0 + \frac{y}{1+\lambda(1-R)^{-1}+\mu} \quad , \tag{27a}$$

$$Z_{eq} = Z_0 + \frac{y}{1+\lambda(1-R)^{-1}+\varepsilon^{-1} m_{star}^{-1}\cdot SFR} \quad . \tag{28a}$$

Equation (29), which is the solution for a steady-state gas-regulated reservoir, is a special case of the more general Equation (8) of Recchi et al. (2008) and Equation (4) of Dayal et al (2012). As noted above, Davé et al (2012) recently considered a model that more closely follows our own approach, and the Equations (25-28) are similar to their Equation (9) except for the important difference that the term representing the flows into, or out of, the gas reservoir in the denominator (i.e. the last term in the denominator given by $\varepsilon^{-1}$ times the bracketed expression in Equations 25-28) is absent. This is because the mass in the gas reservoir does not change with time in their model. In our picture, in which the star-formation rate is *regulated* by the gas reservoir, it is the gas *fraction* $\mu$ that determines the sSFR, as seen from Equation (7). As discussed below, the presence of this term is therefore important as it produces an implicit dependence of the metallicity on the star-formation rate of the system. The three terms in the denominator of the right hand side of Equation (21-24) reflect the three "destinations" of metals as the incoming flow divides, i.e. long-lived stars, removal from the system in an outflow, and the build-up of the gas reservoir.

The metallicity is established instantaneously in the gas-regulated model, because the gas stays only a short time in the system.  The above equations should therefore be valid at any epoch, even if ε and λ are (slowly) varying functions of time, either directly or indirectly through the increasing mass of the system.  In the case of ε, there is some empirical evidence (Daddi et al 2010, Genzel et al 2010) that it is about three times higher at $z \sim 2$ than locally, i.e. that it scales as $(1+z)^{-1}$.  In the absence of empirical evidence to the contrary, we will assume in what follows that the mass-loading factor λ is independent of epoch.

An attractive aspect of the regulator is thus that the metallicity at any point time is set by the current state of the system and not by the past history of it, provided the $d\ln\mu/dt$ term is small (see also Köppen & Edmunds 1999). In other words, the chemical "evolution" of the reservoir is more the changing (and reversible) operation of the regulator than a monotonically increasing temporal development of metallicity due to the build-up of metals.  We will henceforth drop the equilibrium suffix on $Z_{eq}$.

Finally we note that, while we would not expect the nucleosynthetic yield to vary with the mass of the galaxy, the effective yield may depend on mass if the outflowing winds are preferentially enriched relative to the gas reservoir and if this varies with mass.  In the spirit of our heuristic analysis, we will not consider this potential complication.

*3.2  Metallicity relations within the galaxy population*

The analysis in the previous subsection was based on an individual galaxy whose SFR is straight-forwardly regulated, via the gas content, to be a constant (or slowly varying) fraction of the infalling material.  The operation of the regulator is governed by the internal processes parameterized by the SFR efficiency ε and the mass-loading λ of the mass-loss in winds, and the outcome is set by these two parameters plus the external conditions represented by the infall rate onto the system.

If ε and λ are the same for all galaxies of a given stellar mass, and if the variation in the sSFR amongst Main Sequence galaxies (at a given $m_{star}$) reflects long-term smooth variations in their specific infall rates (and not short term stochastic variations), then the Equations (22-25) of the previous section should apply also to the population of galaxies as a whole, allowing us to use the observed metallicity variations within the population (with stellar mass, time and other parameters) to infer the parameters and operation of the regulator system.

Specifically, Equation (26/26a) then establishes a clear linkage between the cosmic evolution of the characteristic sSFR of the Universe as a whole and the cosmic evolution of the metallicity of the stars that are being made (averaged over all systems). Metallicities in star-forming galaxies at earlier epochs will be lower because the sSFRs in stellar systems are generally higher.

Likewise, the explicit appearance of the SFR in Equation (28/28a) allows the possibility of a natural explanation of the claimed $Z(m_{star},SFR)$ relation within the galaxy population at a given epoch, and provides a prediction of how this relation will appear at different epochs. Furthermore, since the $Z(m_{star},SFR)$ relation will only change to the extent that $\varepsilon(m_{star})$ and/or $\lambda(m_{star})$ themselves change with redshift, Equation (28/28a) offers a route to understand the claimed existence of a "fundamental" $Z(m_{star},SFR)$ relation that is independent of epoch (M+10) if these internal parameters of the regulator are indeed more or less constant. A truly epoch-independent FMR is therefore naturally introduced by this model if $\varepsilon(m_{star})$ and/or $\lambda(m_{star})$, which both reflect baryonic processes in galaxies, were independent of epoch since the metallicity of any star-forming galaxy at any-epoch would be given by the single equation of our model, which is given in its different forms as Equations (25-28) with constant parameters. This equation provides the physical basis of the observed stability of the FMR with cosmic (even though we suspect that some small evolution in e is likely, as discussed above in Section 3.1, leading to a perturbation from a strictly constant FMR).

Finally, Equation (29) establishes a very direct link between the observed mass-metallicity relation of galaxies and the mass-dependence of the fraction of baryons that are being converted into stars. We will examine all these different linkages in Sections 4 and 5.

## 4. The dark side of the regulator: links with the dark matter haloes

Under the assumption that Section 3.2 holds, i.e. that the relations derived for an individual gas-regulated galaxy can be used for the overall population, then the Equation (29) directly links the mass-metallicity relation of galaxies to the mass-dependence of $f_{star}$, without requiring direct knowledge of the $\varepsilon$, $\lambda$ or $rsSFR$ parameters which are responsible for setting $f_{star}$ in Equations (12 and 12a). In this section we show that, via $f_{star}$, there is therefore a direct connection between the slope of the mean mass-metallicity relation of galaxies and the relationships between the stellar and dark masses of haloes. To avoid confusion, it should be noted that this ratio of stellar to dark mass is sometimes called, in the cosmological literature, the *star-formation efficiency* of the halo. In this paper, we have however used this term for a quite different quantity (see Equation 4).

In this Section, we focus on the power-law behaviour at low masses. The mass-metallicity relation of T+04 has a logarithmic slope at low masses $m_{star} \sim 10^9 \, M_\odot$ of $\eta \sim 0.35$. The recent stacking analysis of Andrews & Martini (2012) suggests that this may be steeper $\eta \sim 0.5$ at $m_{star} \sim 10^8 \, M_\odot$, and taking a slice through the M+10 $Z(m_{star}, SFR)$ plane along the locus of the Main Sequence also yields a steeper dependence, with $\eta \sim 0.55$ below $10^{10} M_\odot$ (see Figure 9 below).

Provided that the metallicity of inflowing gas $Z_0$ is negligible and if the yield $y$ is independent of the galaxy stellar mass, then Equation (29) can be used to derive the dependence of $f_{star}$ on $m_{star}$. In the mass range where the mass-metallicity relation is a power-law,

$$f_{star} \propto m_{star}^\eta. \tag{30}$$

In the next two sections we explore the implications of this relation between $f_{star}$ and $m_{star}$.

### 4.1 The stellar mass content of dark matter haloes

In considering the stellar mass formed within a given dark matter halo, we need to consider the product of $f_{star}$ and the fraction of baryons which both enter the halo *and* penetrate down to enter the galaxy regulator system, which we have denoted $f_{gal}$. It is the ($f_{star}f_{gal}$) product that gives the incremental build-up of stellar mass in the galaxy relative to the dark matter mass of the halo (always assuming the matter coming into the halo has the cosmic baryon fraction). The final $m_{star}/m_{halo}$ ratio of a given galaxy will be a weighted average of the ($f_{star}f_{gal}$) product that was operating as the galaxy built up its stellar and dark matter components through to the time in question. It is convenient to introduce a further fractional quantity $F$ that represents the integrated stellar to dark mass ratio:

$$F = \frac{\int_0^t f_{star} f_{gal} \frac{dm_{halo}}{dt} dt'}{\int_0^t \frac{dm_{halo}}{dt} dt'} = \frac{m_{star}}{m_{halo}} \ . \tag{31}$$

We will assume, in the power-law regime, a fixed exponent for the ($f_{star}/f_{gal}$) product, so that the mass-dependence of $F$ (=$m_{star}/m_{halo}$) will be the same as the mass-dependence of ($f_{star}/f_{gal}$), i.e. they will have the same logarithmic slope $\eta$ set by the slope of the mass-metallicity relation as in Equation (30).

As discussed in the Introduction, the different faint end slopes of the two mass functions of galaxies and haloes require a strongly varying fraction of baryons in a given halo to be converted into stars,

$$m_{star} \propto m_{halo}^{\gamma}.$$ (32)

with the required $\gamma$ depending on the difference between the faint-end slopes of the Schechter functions of stellar and dark mass respectively, as given by

$$\gamma \sim \frac{1+\alpha_{halo}}{1+\alpha_{stars}}$$ (33)

where we adopt the convention that $\alpha_{halo}$ and $\alpha_{stars}$ are negative. The relevant $\alpha_{stars}$ is that for star-forming galaxies, $\alpha = -1.45\pm0.05$ (Peng et al 2010, Baldry et al 2012), which yields $\gamma \sim 1.9 \pm 0.2$ for $\alpha_{halo} \sim -1.85$ (Guo et al 2010). Rearranging Equation (32) then gives a requirement on the mass-dependence of $F$ (and the $f_{star}/f_{gal}$ product).

$$\eta = \frac{\gamma-1}{\gamma} = 0.46 \pm 0.1.$$ (34)

As noted above, the mass-metallicity relation implies a logarithmic slope $\eta$ of $f_{star}$ with mass in the range $0.3 < \eta < 0.5$ in the mass range $10^8$-$10^{10}$ M$_\odot$ (from Equations 25 and 30). We therefore conclude that the simple gas-regulator model naturally accounts for the variation of the ratio of stellar mass to dark matter mass that is required (Equation 34) to reconcile the faint end slopes of the galaxy and halo mass functions, with a more or less constant $f_{gal}$. In other words, the large variation in stellar to halo mass that is required for the cosmology can arise from baryonic processes operating *within* the galaxy system and not by any significant variation in the fraction $f_{gal}$ of baryons that *enter* the galaxy system (c.f. the discussion in Bouche et al 2010). However, the solution is obviously not unique. If the effective $y$ decreased at low galactic masses because of preferentially enriched winds, $f_{star}(m_{star})$ could be shallower, which would then require a mass-dependence in $f_{gal}$. Our goal is not to rule out more complicated scenarios, only to explore what is possible with the simplest possible model.

As remarked in Section 2, if the fraction $f_{star}$ of incoming baryons that are converted into long-lived stars is more or less constant, then the rsSFR will quickly be set equal to the sMIR$_B$. The corollary is that, if $f_{star}$ is *not* constant, because of changes to the parameters controlling the regulator, then the equality between the rsSFR and the sMIR$_B$ will be perturbed. In particular, if $f_{star}$ for a given evolving galaxy system is *increasing* with time, for example because $f_{star}$ is larger at higher masses because of increased ε (and/or decreased λ) then the rsSFR will be systematically *greater* than the sMIR$_B$. These statements can be extended to the sMIR$_{DM}$ if we consider the ($f_{star}f_{gal}$) product in place of the simple $f_{star}$.

It is again convenient to consider the fractional quantity $F$ introduced in Equation (31). We will then have

$$m_{star} = F \int_0^t \frac{dm_{halo}}{dt'} dt'$$

$$\frac{dm_{star}}{dt} = F \frac{dm_{halo}}{dt} + \frac{dF}{dt} m_{halo} \quad.$$

Dividing by $m_{star}$, we can re-write this in terms of specific quantities, using $m_{star} = F\, m_{halo}$ from the definition in Equation (31),

$$rsSFR = sMIR_{DM} + \frac{1}{F} \frac{dF}{dt}. \tag{35}$$

The second term on the right is the boost to the rsSFR that comes from the change in $F$ as the galaxy grows. This could be set to zero in the discussion of the "ideal" regulator in Section 2.1, but in the more general case that we are considering, it will have two components: the change in $F$ with stellar mass as the stellar mass of the galaxy increases, plus any temporal change of $F$ at fixed mass

$$rsSFR = sMIR_{DM} + \frac{1}{F} \frac{\partial F}{\partial m_{star}} \frac{\partial m_{star}}{\partial t} + \frac{1}{F} \frac{\partial F}{\partial t} \tag{36}$$

If $F$ and $f_{star}$ have the same mass dependence (as in the previous subsection), then the middle term is simply the product η·$rsSFR$, where η is the logarithmic slope of the $f_{star}(m_{star})$ relation in Equation (30). We will neglect the last term for the time being, since it should be small, returning to this in Section 5 below.   We then get by simple rearrangement

$$rsSFR \sim \frac{1}{1-\eta} sMIR_{DM} \quad . \tag{37}$$

With our observed value of $0.35 < \eta < 0.55$ below $m_{star} \sim 10^{10} \, M_\odot$ from the mass-metallicity relation, we would expect that the observed rsSFR would be boosted by a factor of roughly 0.2 to 0.35 dex relative to the $sMIR_{DM}$.   This is about right to explain the comparison between the observed rsSFR and the $sMIR_{DM}$ in Figure 1, especially at low redshifts ($z < 1$) where the offset is about 0.3 dex.  The offset is apparently larger at higher redshifts, possibly being as high as 0.6 dex at $z \sim 2$.

Furthermore, if the logarithmic slope $\eta$ flattens with increasing mass, as indicated by the curvature of the mass-metallicity relation, then this boost factor would be lower at high masses and this will change the logarithmic slope of the rsSFR($m$) and $sMIR_{DM}(m)$ relations, offering a qualitative explanation of the opposite signs of the $\beta$ of galaxies and haloes.  We will return to this question in a more quantitative way in Section 5.

This explanation of the observed rsSFR > $sMIR_{DM}$ is distinct from the explanation offered by Bouche et al (2010) in terms of large variations in $f_{gal}$.  As in the previous section, our own analysis indicates that the variation in $f_{star}$ with mass *inside* the galaxy system is sufficient explanation and that $f_{gal}$ can instead be more or less constant with mass.  Again our goal is not to rule out the Bouche et al threshold scheme, only to suggest that it may not be required.

## 5.      Fitting the observed $Z(m_{star}, SFR)$ and $Z(m_{star})$ relations

As noted in the Introduction, Ellison et al (2008), Mannucci et al (2010, M+10), Lara-Lopez et al. (2010) and most recently Andrews et al (2012), have all drawn attention to evidence that the SFR of galaxies may act as a second parameter in the mass-metallicity relation.  M+10 moreover suggested that the $Z(m_{star}, SFR)$ relation is invariant with epoch out to redshifts $z > 2$, coining the phrase "fundamental metallicity relation", or FMR.

In the light of a number of difficulties in determining gas-phase metallicities (see e.g. Yates et al 2012, Andrews & Martini 2012 for discussion), there is an ongoing debate about the form of the $Z(m_{star}, SFR)$ relation and whether it is truly independent of epoch.  Different metallicity estimators are often combined, some of which are known to have saturation issues at high metallicities (e.g. Kewley & Ellison 2008, Lara-Lopez et al 2012). The use of the Hα line in both metallicity and SFR measurements may introduce coupling of errors, and potential correlations between SFR and ionization parameters may also be present. Also, in samples spanning a significant redshift range, SFR and epoch may be coupled due to the cosmic evolution of the sSFR. As a result of these and other issues, the validity and form of the $Z(m_{star}, SFR, z)$ relation are still being debated.

With this caveat in mind, it is nevertheless interesting that, from Equation (28), a simple $Z(m_{star}, SFR)$ relation linking metallicity to stellar mass and star-formation rate is a natural outcome of the simple operation of the regulator model. We would also expect that the SFR dependence of the mass-metallicity relation would be stronger at lower masses if the star-formation efficiency was lower at lower galactic masses. Furthermore, the $Z(m_{star}, SFR)$ relation would also be epoch-invariant in our model, if $\varepsilon(m_{star})$ and $\lambda(m_{star})$ did not change with redshift. As discussed in Section 3.1, we would in fact expect some change in $\varepsilon(m_{star})$ from observations of the gas depletion timescales, but this change is much smaller (a factor of 3 to $z \sim 2$) than the change in SFR at fixed mass, which is a factor of 20 to the same redshift.

### 5.1 Fitting the M+10 Z(m_{star}, SFR) SDSS data

We therefore compare our Equation (28) to the tabulated $Z(m_{star}, SFR)$ data for SDSS galaxies that have been presented by M+10 in their Table 1. Since we expect both $\varepsilon$ and $\lambda$ to vary with stellar mass, this comparison amounts to fitting the data with Equation (28) with $\varepsilon(m_{star})$ and $\lambda(m_{star})$ as free functions. In our fits, we will assume that both $\varepsilon$ and $\lambda$ may be represented by power-laws in the stellar mass of the galaxy, i.e.

$$\lambda = \lambda_{10} m_{10}^a$$
$$\varepsilon = \varepsilon_{10} m_{10}^b. \tag{38}$$

with $m_{10}$ the stellar mass in units of $10^{10} M_\odot$. We will make the heuristic assumption that $y$ and $Z_0$ are not functions of galactic mass.

We now return to evaluate the $d\ln\mu/dt$ term discussed in Sections 2 and 3. Since $\varepsilon$ scales (with our assumed time dependence) as $m_{star}^b t^{-1}$ and the sSFR as $m_{star}^g t^{-2.2}$ it is easy to see, given $\mu = \varepsilon^{-1} \cdot sSFR$, that

$$\frac{d\ln\mu}{dt} = -\frac{1.2}{t} + (\beta - b)rsSFR \qquad (39)$$

When multiplied by $\varepsilon^{-1}$, this has a numerical value around -0.25, i.e. "small" compared with the other terms of order unity, but not negligible. With this, Equation (28) may be written

$$Z_{eq} = Z_0 + \frac{y}{1 + \lambda(1-R)^{-1} + \varepsilon^{-1}\left((1+\beta-b)m_{star}^{-1} \cdot SFR + (1-R)^{-1}\frac{1.2}{t}\right)} . \qquad (40)$$

The addition of the two components of the $d\ln\mu/dt$ term acts to introduce an additional constant term in the denominator (since $\varepsilon$ is expected to scale roughly as $t$) and to weaken the dependence on the SFR by a factor $(1+\beta-b)$. This is the equation that can now be fit to the SDSS data, setting $t \sim 13.8$ Gyr, so the last term in the denominator takes a value $-0.15$. Equation (40) can also be used to predict the evolution of the $Z(m_{star},SFR)$ relation. We then have five free parameters, $\varepsilon_{10},\lambda_{10},a,b,y$ in Equation (28), or six if the infall metallicity $Z_0$ is allowed to vary.

The fits are done using a $\chi^2$ statistic, using the dispersion of $Z$ in a given ($m_{star},SFR$) bin divided by the square root of the number of galaxies (both taken from Table 1 in M+10) as the uncertainty at each point. The best-fit parameters are given in the upper part of Table 1. The quoted uncertainties in each quantity reflect only the nominal uncertainties in the parameter estimation that come from the $\chi^2$ analysis. A better idea of realistic systematic uncertainties comes from examining the range of values in the Table. The three fits are with the infall metallicity $Z_0$ constrained to have three values relative to the yield $y$.

Figure 5 shows that Equation (28) is well able to reproduce the observed M+10 $Z(m_{star},SFR)$ surface. The r.m.s. deviations in $Z$ across the ($m_{star},SFR$) plane are about $0.015-0.018$ dex for the different fits (see Figure 5), compared with a typical dispersion in $Z$ within the population at fixed ($m_{star},SFR$) of about 0.07. It was found that fits which were not weighted by the number of galaxies in the bin gave almost identical values of the parameters.

The returned values for $\varepsilon(m_{star})$ and $\lambda(m_{star})$ are quite reasonable from independent astrophysical considerations. The fitted $\varepsilon_{10}$ SFR efficiency at $10^{10}M_\odot$ corresponds to a gas depletion timescale $\tau_{gas}$ of 2-3 Gyr for $0 < Z_0/y < 0.1$. This is longer than measurements of the consumption timescale of *molecular* gas at the present epoch (Young and Scoville (1991), Daddi et al (2010), Genzel et al 2010, Saintonge et al 2011b, and references therein), which are of order 1-2 Gyr, but additional atomic gas should be included, since it will take part in the chemical mixing of the galaxy.  This is typically comparable in mass (e.g. Young & Knezek 1989, Saintonge et al 2011a) and its inclusion will therefore roughly double $\tau_{gas}$.  As the only term in Equation (28) that includes the SFR, $\varepsilon$ will have been determined by the variation in $Z$ with SFR at fixed mass in the M+10 data.  However, as discussed above, it's numerical value will have been affected by the value of $b$, as in Equation (40), and $\varepsilon$ would be higher (and $\tau_{gas}$ lower) if $b < 0.3$.

Turning to the outflows, the best fit $\lambda_{10}$ from the fits to the M+10 data is in the range $0.2 < \lambda < 0.3$, which is quite low compared with estimates of galaxies at moderate (Wiener et al. 2005) and high (Newman et al., 2012) redshifts.   Furthermore, there is a rather strong inverse dependence of $\lambda$ on stellar mass, with $a \sim -0.9$.  Theoretical models generally produce a somewhat weaker inverse relationship with mass, e.g. the momentum-driven wind model of Murray et al (2005) has $\lambda \propto m^{-1/3}$, whereas an energy-driven wind (e.g. Dekel and Silk 1986) has $\lambda \propto m^{-2/3}$.  Observationally, the mass dependence of outflows is not well constrained. Observations of outflowing entrained MgII material both in the "down-the-barrel" spectra of star-forming galaxies and as seen against background galaxies at projected distances out to 40 kpc both imply a significant inverse dependence on mass (see Weiner et al 2005, Bordoloi et al 2011).  Newman et al (2012) have suggested little mass dependence on $\lambda$ in star-forming galaxies at $z \sim 2$, above a strong threshold in surface SFR density.  The steep dependence on stellar mass of the wind mass loading $\lambda$ in the fits can be traced to the strong curvature of the overall $m$-$Z$ relation in M+10 and in particular to the flattening at high masses, which requires that winds become negligible at these masses.   We also note that introducing a mass-dependent yield or inflow $Z_0$ would change the required $\lambda(m_{star})$ dependence.

Finally, the value of the yield $y$ in Table 1 is quite high, about +9.0 in units of 12+log(O/H), i.e. about $y = 0.016$ as a mass ratio. This can be compared with e.g. 0.004 in the analysis of Dalcanton et al (2007).  This high value is not however outside of the range of theoretical values for a Salpeter initial mass function (see the compilation in Table 2 of Henry et al 2000, correcting for the mass return fraction which is not included there). Furthermore, the value of $y$ is driven by the overall normalization of the mass-metallicity relation. This is empirically uncertain by 0.4 dex (a factor of 2.5) as seen on Figure 2 of Kewley & Ellison (2008) and the metallicities of star-forming galaxies used here, which go up to 12+log(O/H) = 9.1 lie at the upper end of the range in the literature.

Overall, the parameters returned from the fits are not unreasonable, and it is therefore interesting that the predicted $Z(m_{stars}, SFR)$ relation from Equation (28) appears to be able to reproduce the M+10 data.  We stress again that the point of this exercise is not to try to determine observationally the values of ε, λ, or *y*.  Clearly, even within the framework of the model, there are large systematic uncertainties driven by our choices of constant $Z_0$ and *y*. Furthermore, as commented earlier, the form of the empirical $Z(m_{star}, SFR)$ relation in real galaxies is by no means settled.

### 5.2  Fitting the Tremonti et al (2004) $Z(m_{star})$ relation

To explore the effects of systematic uncertainties in metallicity measurements, we also fit using the same Equation (40) to the $Z(m_{star})$ mass-metallicity relation of Tremonti et al. (2004, T+04).  This shows less curvature and a flatter low mass slope than the M+10 data.  To fit the $Z(m_{star})$ relation we must impose a value of ε and also apply a mean SFR-mass relation for Main Sequence galaxies, using for this purpose Equation (1) of Peng et al. 2010.  For ε, we impose the ε($m_{star}$) function from our fits to the M+10 data from Section 5.1, since we argued above it is more or less consistent with observational estimates of $\tau_{gas}$, as discussed above.  Setting ε($m_{star}$)  as indicated in the second half of Table 1, yields a weaker mass-dependence for λ when we fit the Tremonti et al (2004) mass-metallicity relation, with *a* ~ −0.5, as shown in the second part of Table 1, and a higher level at high masses, with $\lambda_{10}$ ~ 0.5.

Clearly, the systematic uncertainties in gas-phase metallicities are a significant limitation at the present time. Not least, the numerical fits depend sensitively on the slope and curvature of the mass-metallicity relation.  It will be important to return to this kind of analysis when the observational uncertainties have been substantially reduced.   This, coupled with the simplicity of the model, means that the values of ε($m_{star}$) and λ($m_{star}$) cannot be considered "measurements".  Nevertheless, we will explore in the next sub-sections the implications of both sets of fits that we obtained from this Section, taking the various estimates of  ε($m_{star}$) and λ($m_{star}$) in Table 1 at face value.  Finally we note that the approach adopted here is quite similar to that taken by Peeples & Shankar (2011) who used a close analogue of Equation (27), together with observational estimates of the gas ratio μ to fit the mass-metallicity relation.

### 5.3   The destinations of baryons at high and low redshift

We compute the fractions $f_{star}$, $f_{res}$, and $f_{out}$ using Equations (26-28). These are shown in Figure 6 for the fits to the M+10 $Z(m_{star}, SFR)$ data from Section 5.1, and for the constrained fits to the T+04 $Z(m_{star})$ data described in Section 5.2.   As discussed earlier, ε is assumed to scale as (1+z) but λ is taken to be independent of epoch.

Mass loss from the system dominates at low masses and the formation of stars dominates at high masses.  At high redshifts, where the gas fractions will be substantially higher because of the higher value of the (ε⁻¹·*sSFR*) product, the flow of gas into the reservoir is the dominant destination of baryons in intermediate mass galaxies, highlighting the potential importance of this term (which, as noted above, is omitted in some previous analyses, e.g. Davé et al 2012). At low redshifts, the flow into the reservoir is negative for all galaxies, i.e. the reservoirs are gradually depleting their gas at the present epoch.  This must be the case (for more or less constant ε) if the SFR in a given galaxy is declining.  However that this net rate of decrease of the gas in the reservoir is still a small fraction of the continuing infall Φ (which has unit strength relative to the values of $f_{star}$, $f_{res}$, and $f_{out}$ shown in Figure 6), i.e. of the gas flow *through* the reservoir.

*5.4  Prediction for the mean mass-metallicity relation at high redshift*

Figure 7 shows the predicted mean $Z(m_{star})$ relation for galaxies on the Main Sequence at high redshifts using the parameters for the $0 < Z_0/y < 0.1$ fits at $z \sim 0$ and including an assumed (1+z) dependence of ε, using the cosmic evolution of the sSFR from Equation (2).   The evolution in $Z(m_{star})$, it should be recalled, come about *entirely* through the (small) observed changes in the star-formation efficiency ε and the much larger observed increase in the sSFR in Equation (28) so there are in principle no free parameters in deriving these predictions. Having said that, the change in sSFR is much better determined than the change in ε, and it is the product of these, i.e. μ, that determines the metallicities.  To illustrate this, we also show in Figure 7 also the case where ε is held constant, i.e. producing an absolutely constant FMR. As would be expected, this produces larger changes in metallicity.  It should be noted that the evolution in the $Z(m_{star})$ relation with evolving ε slows down at high redshifts as the (1+z) dependence of ε largely nullifies the $(1+z)^{5/3}$ increase in sSFR.

The $z \sim 2$ data of Erb et al (2006) are over-plotted on Figure 7, to be compared with the highlighted $z \sim 2$ prediction. These data are based on NII/Hα and so cannot be directly compared to the model which is based on the SDSS [O/H] data at low redshifts.  Furthermore, NII/Hα has serious saturation effects at high metallicities, relevant for massive $m_{star} > 10^{10} M_\odot$ galaxies at low redshift (see Erb et al 2004).  On both diagrams we have converted the NII-derived metallicities to the T+04 scheme applying the conversion equation given by Kewley and Ellison (2008). This not very satisfactory procedure highlights the importance of obtaining fully consistent metallicity estimates of galaxies over a wide range of redshift. Mindful of this caveat, our simple model clearly reproduces qualitatively well the changes with redshift in the overall mass-metallicity relation, independent of the detailed fits to the low redshift data.

The success of the gas-regulated model in reproducing the temporal evolution of the gas metallicities of galaxies and the mass- and SFR-dependence at a given epoch suggests that these may both be viewed as manifestations of the same basic operation of the gas-regulation process in galaxies, operating at all epochs. In this view, there is only one equation (in its different forms given by Equations 26-28) determining the metallicity of a galaxy, regardless of epoch. This ultimately comes about because of the short $\tau_{gas}$ compared with the timescales on which the external conditions, and internal parameters of the regulator, are changing. The consequent rapid flushing of gas through the system ensures that the metallicity of the gas is an instantaneous reflection of the "state" of the system, i.e. the relative importance of the different destinations of the baryons. In this sense, there is no chemical "evolution" of galaxies *per se*, merely the slowly changing instantaneous operation of the regulator. This is evident in Equation (29) where, neglecting the $Z_0$ of the infalling material, the gas metallicity is set by the instantaneous $f_{star}$ (see also Davé et al 2012), which in turn is set (neglecting the small $d\ln\mu/dt$ term) by the instantaneous values of $\varepsilon$, $\lambda$ and sSFR.

*5.5 The results of Section 4 revisited*

Finally, we can use the fitted values of $\varepsilon(m_{star})$ and $\lambda(m_{star})$ given in Table 1 to construct a simple numerical model of evolving galaxies within their dark matter haloes. Specifically we use the six fits in Table 1 with $0.0 < Z_0/y < 0.1$, and include the observed $(1+z)$ redshift dependence of the $\varepsilon$ parameter. This numerical model enables us to further explore and validate the analytic results linking the stellar and dark matter mass that were derived in Section 4, i.e. the $m_{star}$ dependence of the ratio $m_{star}/m_{halo}$, and the offset between the rsSFR and the sMIR$_{DM}$, outside of the power-law regime that was explored in the analytic analysis.

We construct the numerical model as follows. A large number of haloes of different initial mass grow according to Equation (3). Starting at an arbitrarily early point (50 million years after the Big Bang) gas is fed in to each galaxy system at a rate $\Phi$ that is set to be a constant multiple of the rate of baryonic mass increase of the associated halo, assuming a baryonic fraction of 0.15. Gas in the galaxy is then continuously transformed into stars, and ejected from the system in winds, according to the parameters $\varepsilon(m_{star}, z)$ and $\lambda(m_{star})$ that are listed in Table 1. The constant multiplicative factor ($f_{gal} = 0.4$) is chosen so that the simulation results, at the present epoch, in a galaxy with $m_{star} = 10^{10.5}$ $M_\odot$ in a $10^{12}$ $M_\odot$ mass halo (e.g. Guo et al 2010).

Figure 8 shows the ratio $m_{star}/m_{halo}$ that is produced by this numerical model, both as a function of epoch at fixed (observed) $m_{star} = 10^{10} M_\odot$ (in the left hand panel) and as a function of $m_{star}$ at $z = 0$ (in the right right hand panel). The dashed line in the right hand panel show the relation ($\eta \sim 0.45$) that is required to match the faint end slopes of the galaxy and halo mass functions, as discussed in Section 4.1. It can be seen that, as expected from the earlier analytic discussion, the numerical model reproduces quite well the required $m_{star}$-dependence of $m_{star}/m_{halo}$. The fit to the SDSS M+10 data does a bit better because it has a steeper mass-metallicity relation than the T+04 sample. The $m_{star}/m_{halo}$ ratio at fixed $m_{star}$ weakly declines with redshift.

Figure 9, which is based on Figure 1 in the Introduction, shows the rsSFR output from the numerical model, again (left hand panel) as a function of epoch at (observed) stellar mass of $10^{10}$ $M_\odot$, and (on the right hand panel) as a function of stellar mass at both $z = 0$ and $z = 2$, compared with the observed values (in red) that were summarized in Equation (2) and plotted in Figure 1. The increase in the rsSFR, relative to the dark matter $sMIR_{DM}$ is clearly seen and has a value (more or less independent of redshift) of about 0.35 dex for the fits to the M+10 data (in blue) and 0.25 dex for the fits to the T+04 data. These boosts of the rsSFR relative to the $SMIR_{DM}$ are very close to the numbers expected from the analytic Equation (37).

It can also be seen, in the right hand panel of Figure 9, that the mass-dependence of the rsSFR is reversed from that of the $sMIR_{DM}$ of the haloes. This is because of the curvature in the $f_{star}(m_{star})$ relation that can itself be traced to the curvature in the $Z(m_{star})$ mass-metallicity relation. The output from numerical model well matches the $\beta \sim -0.1$ slope of the observed SDSS rsSFR($m_{star}$) relation (Peng et al 2010).

It should be noted, as remarked earlier in the paper, that the numerical model, using the $\varepsilon$ and $\lambda$ values from the fits to the metallicity data, has successfully reproduced the mass dependence of the rsSFR with an $f_{gal}$ that is independent of mass. Taken at face value, the feeding of galaxies from the halo should not be strongly dependent on the mass of the galaxy. The mass-dependence of $m_{star}/m_{halo}$ comes from the *internal* action of the regulator system, as parameterized by $\varepsilon$ and $\lambda$ and as seen observationally in the mass-metallicity relation. Linked to this, the (constant) value of $f_{gal} \sim 0.4$ that was required to match the normalization of the $m_{star}$-$m_{halo}$ relation, i.e. $m_{star}/m_{halo} \sim 0.03$ at $m_{star} \sim 10^{10} M_\odot$, implies that a significant fraction of baryons that enter a halo penetrate down to be cycled through the galaxy system, even though, at low masses, most of them are subsequently expelled back out of the galaxy system into the halo (or beyond).

It is apparent on both panels of Figure 9 that the boost of the rsSFR relative to the sMIR$_{DM}$ is not quite enough at high redshifts, $1 < z < 3$. Our model, especially with the parameters derived from the M+10 data, has closed at $z \sim 2$ about half the gap between the observed rsSFR and the sMIR$_{DM}$ (from simulations), but a deficit of about 0.3 dex still remains. This apparent deficit is comparable to systematic uncertainties in measurements of star-formation rates, especially when systematic uncertainties like the initial mass function are included. If real, this shortfall might indicate that $f_{gal}$ has an epoch-dependence, producing an additional boost through the time-derivative term in Equation (36).

## 6. Summary and Conclusions

In this paper, we have explored the operation of a simple gas-regulated model of galaxies in order to gain insights into the evolution of galaxies. The main thrust has been to establish links between, on the one hand, the growth of galaxies, specifically the cosmic evolution of the sSFR and the specific accretion rate of dark matter haloes, and, on the other hand, the metallicity of the evolving population of galaxies. The model differs from other similar presentations by including the effects of the varying gas reservoir, which acts to regulate the SFR in the galaxies. The main points of the paper may be summarized as follows.

1.  The simple gas-regulated model explored here assumes, for a given galaxy, that (a) the star-formation rate is proportional to the mass of gas present (SFR=$\varepsilon^{-1} \cdot m_{gas}$), and (b) that any wind outflow is proportional to the star-formation rate ($\Psi = \lambda \cdot$SFR). These are heuristic assumptions adopted to make the model analytically simple.

2.  In its most ideal (and unrealistic) situation in which the parameters describing the star-formation efficiency $\varepsilon$ and the mass-loading $\lambda$ of the wind are constant, the regulator has the key feature of setting the reduced specific star-formation rate (rsSFR) to be equal to the specific baryonic infall rate of the galaxy, sMIR$_B$, which is defined in Equation (16) as the infall rate divided by the past time integral of the infall rate. Although this is unobservable itself, it will likely be closely linked to the specific mass increase rate of the halo (sMIR$_{DM}$) and will be identical if a constant fraction $f_{gal}$ of mass entering the halo penetrates down to enter the galaxy system as gas. More realistic systems with evolving ε and λ will perturb this identity but preserve a close connection between the $sMIR_{DM}$ and the rsSFR. Aside from its attractive physical simplicity, the gas-regulated model is therefore motivated by the broad similarities of the observed rsSFR of galaxies on the Main Sequence and the typical $sMIR_{DM}$ "observed" in cosmological numerical simulations over a wide redshift redshift range from $z \sim 0$ up to at least $z \sim 2$.

3. The metallicity of the gas-reservoir in a simple gas-regulated galaxy is set "instantaneously" by the constant or slowly varying parameters of the regulator, i.e. the SFR efficiency ε, the wind mass-loading λ, and the sSFR (or $sMIR_B$). The metallicity is largely independent of the evolutionary path that the galaxy has been followed hitherto. This is because gas is continuously flushing through the system, since the gas consumption timescale $\tau_{gas}$ is short.

4. If the star-formation efficiencies and wind mass-loading parameters are similar across the galaxy population at a given stellar mass, and if the observed variation in the sSFR of Main Sequence galaxies (again, at a given stellar mass) reflects a variation in the (slowly varying) inflow rates onto galaxies, then the relations derived for an individual system will also apply to the population of galaxies, both at a given epoch and also across time.

5. The gas-regulated model naturally produces an implicit dependence of metallicity on the SFR, as given by Equation (28). This arises because the relative size of the gas reservoir is linked to the sSFR in the gas-regulated model. There is also a dependence of metallicity on the mass of the galaxy, especially if ε and λ vary with mass. The model therefore naturally produces a $Z(m_{star}, SFR)$ relation.

6. Furthermore, the $Z(m_{star}, SFR)$ relation will only evolve with redshift to the extent that the parameters ε and λ of the regulator themselves change (at fixed stellar mass) with epoch. An epoch-independent "fundamental metallicity relation" would be expected if ε and λ, which both reflect baryonic processes within galaxies, are the same at different epochs, and the model gives the physical basis for why the $Z(m_{star}, SFR)$ relation should be stable over cosmic time.

7. There is moreover a direct link between the instantaneous metallicity and the fraction $f_{star}$ of baryons that enter the system and are transformed into long-lived stars, without requiring knowledge of the particular values of ε, λ or the sSFR.

8. The link between metallicity and $f_{star}$ can therefore be used to establish the (stellar) mass dependence of $f_{star}$ from the observed mass-metallicity relation. The implied steep dependence $f_{star}$ on mass reconciles the different faint end slopes of the galaxy and halo mass functions and does so with a more or less constant $f_{gal}$, the fraction of baryons entering the halo that penetrate down to enter the galaxy system.

9. The strong dependence of $f_{star}$ on galactic mass also implies that a given galaxy will be increasing its $f_{star}$ as it grows in mass. This has the consequence that the rsSFR will be elevated relative to the sMIR$_B$ (and thus the sMIR$_{DM}$ if $f_{gal}$ is constant). The boost will be bigger at lower masses. This effect naturally explains why the observed Main Sequence rsSFR is systematically higher than the sMIR$_{DM}$ derived in dark matter simulations, and why the mass-dependences of the rsSFR and sMIR$_{DM}$ are reversed.

10. With a cautionary note that there is an open debate on the validity of SFR as a second parameter in the galaxy mass-metallicity relation, and on whether there is a universal epoch-independent $Z(m_{star}, SFR)$ relation, we can fit the SDSS $Z(m_{star}, SFR)$ data of Mannucci et al. (2010) with our predicted $Z(m_{star}, SFR)$ relation given in Equation (28). The fitted $\varepsilon(m_{star})$ returns a sensible gas depletion timescale consistent with observations. The returned values of the mass-loading of the wind $\lambda$ are also not unreasonable, although they have quite a steep mass dependence that may be traced to the strong curvature of the mean mass-metallicity relation in the M+10 data. Fitting the $Z(m_{star})$ mass-metallicity relation of T+04, which requires us to assume a form for $\varepsilon(m_{star})$ and to impose a mean SFR-$m_{star}$ relation, yields a higher and less mass-dependent form for $\lambda$. Future observational progress on both the form of the $Z(m_{star}, SFR)$ relation and on the mass-loading of winds will be helpful.

11. Taking at face value the parameters $\varepsilon(m_{star})$ and $\lambda(m_{star})$ at zero redshift from both sets of fits and imposing an observationally motivated $\varepsilon \propto (1+z)$ the model also qualitatively reproduces the evolution in the mass-metallicity relation to $z \sim 2$. The small change in $\varepsilon$ (which actually *increases* the metallicities at high redshift) is more than compensated by the much larger decrease that is associated with the much higher sSFR in galaxies at $z \sim 2$, since it is the $\varepsilon^{-1}sSFR$ product that counts.

12. Finally, a simple numerical model in which gas is fed into a galaxy at a rate proportional to the increase in mass of the dark matter halo, and in which star-formation and mass-loss are governed by the $\varepsilon(m_{star})$ and $\lambda(m_{star})$ returned by the fits, verifies that the run of $m_{star}$ with halo mass, and the boosting of the rsSFR relative to the sMIR$_{DM}$, are as observed in the sky, with the possible exception that the boost at high redshift $z \sim 2$ may not be quite large enough.

This paper therefore establishes a direct linkage between the global evolution of the sSFR in the Universe and the metallicities of the stars that are formed throughout that evolution, and also between the sSFR of galaxies and the growth of their dark matter haloes. It also establishes a direct link between the ratio of stellar to dark mass in galaxies (required to reconcile the faint end slopes of the galaxy and halo mass functions) and the observed enhancement of the rsSFR relative to the halo mass increase rates. All of these connections are achieved through the action of the single, very simple, gas-regulator system, which acts on the inflow of gas onto galaxies, splitting it into three branches: baryon-storage in long-lived stars, ejection from the system in a wind, and flow into or out of the gas reservoir that regulates the star-formation.

One of the striking things in the analysis is how the linkages made above with the dark matter haloes arise because of baryonic processes operating *within* the galaxies. The relations with dark mass have the correct form only if $f_{gal}$, which we introduced as the fraction of baryons that enter the halo and penetrate down to enter the galaxy system itself, is more or less independent of mass. Furthermore, it was shown that $f_{gal} \sim 0.4$ was required to yield the correct $m_{star}/m_{halo} \sim 0.03$ value at $m_{star} \sim 10^{10.5} M_{\odot}$. Taken at face value, this implies that of order a half of the baryons entering a halo are cycled through the galaxy system, even if many of them are ejected again in winds. Baryonic processes operating within the regulator system, and dependent on the stellar mass of the galaxy, are largely responsible for the variation in stellar mass with halo mass, rather than variations in the fuelling of galaxies in different mass haloes.

The linking of the specific growth rates of stars and dark matter that is produced by the regulator, together with Equation (7), may lead to a number of new perspectives on the evolution of galaxies. As an example, we can argue that high redshift galaxies *must* be gas-rich *because* they must have a high sSFR (because their haloes have a high $sMIR_{DM}$) rather than the other way around.

One of the novel things in the paper has been the natural emergence of SFR (or gas ratio) as a second parameter in the mass-metallicity relation of galaxies. The fact that the fit of Equation (28) to the $Z(m_{star}, SFR)$ data that have been presented by Mannucci et al (2010) successfully returns a reasonable value of the gas consumption timescales in galaxies $\epsilon^{-1}$ (at least when the infall $Z_0$ is low) suggests that the spread in SFR in the Main Sequence population at fixed mass, which drives the returned value of $\epsilon$ in the fits, arises from long-term variations in the infall rate of material onto the galaxies.

Finally, we can look at the role of "feedback" processes in the gas-regulated model. Feedback has often been discussed in the context of star-formation in galaxies and the wind outflow, given by $\lambda \, SFR$, is in a sense "feedback" from star-formation. However, as discussed in the Paper, this wind has no bearing at all on the strong link between the *sSFR* and the *sMIR_B,* nor on the gas ratio μ within the galaxy, which is set only by ε and the sSFR. We have also argued that the equations governing the behavior of *individual* galaxies are also applicable to the galaxy population as a whole, as evidenced by the success in reproducing the mean $Z(m_{star}, SFR)$ relation and by the modest scatter of the data around this relation (typically 0.07 dex in *Z*). This implies the parameters ε and λ must be quite uniform across the Main Sequence population (at a given epoch and a given stellar mass) and are not affected by events in individual galaxies.

We have repeatedly used the term "simple model" in this Paper, to emphasize the simplifying assumptions on which it has been based. Not least the model assumes homogeneous mixing of the gas within the system and continuous flushing of the system with incoming gas. There is a hard boundary with the outside, and outflowing material is assumed to be lost forever. The metallicity of infalling material is taken to be constant and largely negligible. We have assumed constant yield, i.e. that the wind outflow is not preferentially enriched as a function of galactic mass. We have neglected the possible mixing of the outflowing enriched material with the gas in the halo. With these significant caveats in mind, the analysis nevertheless represents a good starting point for considering the chemical evolution of galaxies over a broad range of cosmic time. The evolution of the gas metallicity will, of course, be linked to the development of the metallicities of the stars in a galaxy and this will be explored in a later paper.

The model should not be taken as a precise quantitative model for galaxies nor can it be used to exclude other more complex scenarios. Nevertheless, the success of the current analysis suggests that the underlying gas-regulated model of galaxy evolution has some considerable validity as a basic description of the galaxy population over a wide range of epochs and further suggests that it may be possible to view the chemical abundances in galaxies as arising instantaneously from the operation of this basic regulatory system. The model provides an excellent starting point for considering the development of the stellar populations and metal content of galaxies in the context of their dark matter haloes.

**Acknowledgements**

This work has been supported by the Swiss National Science Foundation. We thank Nicolas Bouche for stimulating discussions at several points in the development of this work, Christian

**Appendix A**

Let $Z'$ be the deviation from the steady state metallicity $Z_{eq}$ given by equation (26-28), i.e.

$$Z' = Z - Z_{eq}. \tag{A1}$$

We then substitute $Z$ from A1 into equation (24) with the last term set to zero, and use Equation (7) o express $\varepsilon^{-1} rsSFR$ in terms of $\mu(1-R)$,

$$\varepsilon^{-1} \frac{dZ'}{dt} = y(1-R) - (Z' + Z_{eq} - Z_0)((1-R)(1+\mu) + \lambda). \tag{A2}$$

Then using equation (25) to eliminate $Z_{eq}$-$Z_0$ and thus $y$, we have

$$\varepsilon^{-1} \frac{dZ'}{dt} = -Z'((1-R)(1+\mu) + \lambda), \tag{A3}$$

or

$$\frac{d\ln Z'}{dt} = -((1-R)(1+\mu) + \lambda) \ \tau_{gas}^{-1}. \tag{A4}$$

The metallicity of the gas reservoir therefore approaches the equilibrium value exponentially with a timescale of order the gas consumption timescale. The actual timescale to approach equilibrium is given by

$$\frac{\tau_{gas}}{((1-R)(1+\mu)+\lambda)} = \frac{SFR}{\Phi}\tau_{gas} \leq \tau_{gas}. \tag{A5}$$

Figure 1: Comparison of the observed rsSFR of galaxies (red points taken from the compilation of Stark et al 2012, see references in text), summarizd by Equation (2) (red lines in both panels), and the specific mass increase rate of dark matter haloes (sMIR$_{DM}$) in numerical simulations (black lines in both panels) from Equation (3). In the left panel, the rsSFR is plotted for a $10^{10}M_\odot$ galaxy and for a $10^{11.5}M_\odot$ halo at different epochs. The sSFR is systematically about a factor of two higher than the sMIR$_{DM}$, but the evolution with redshift is very similar. In the right hand panel, the rsSFR and sMIR$_{DM}$ are shown as a function of mass at $z \sim 2$ and $z \sim 0$ (lower and upper axes respectively). The rsSFR has logarithmic slope $\beta \sim -0.1$, whereas the sMIR has $\beta \sim +0.1$.

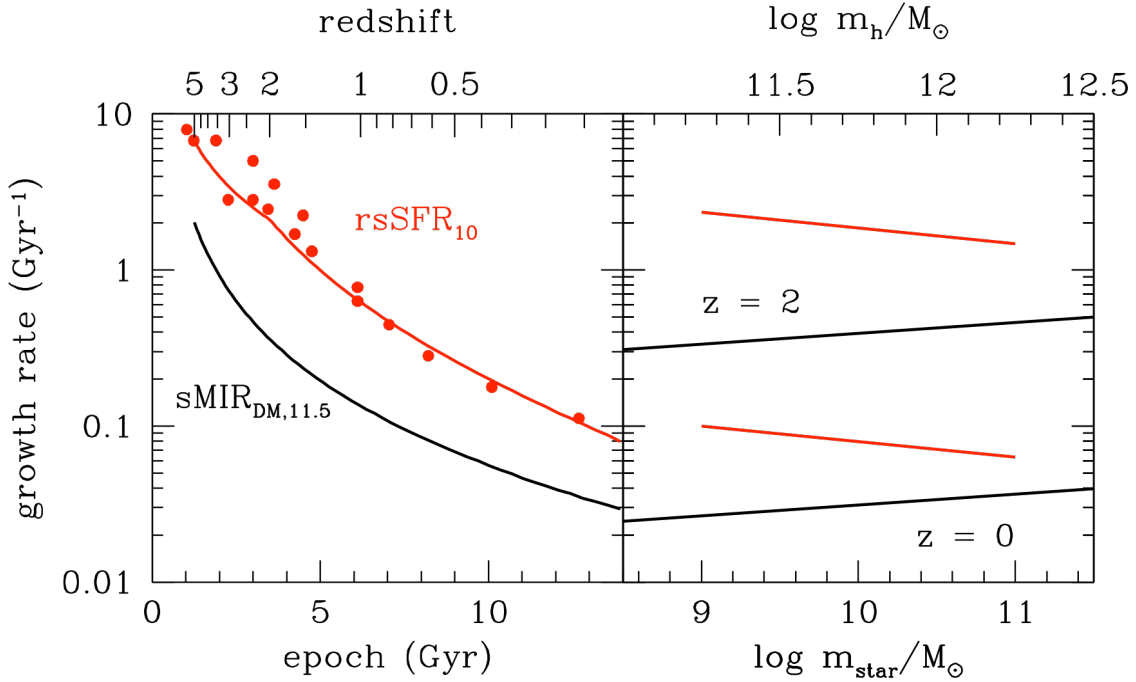

Figure 2: Illustration of the gas-regulated model, in which the SFR is regulated by the mass of gas in a reservoir within the galaxy. Gas flows in to the halo, some fraction $f_{gal}$ of which also flows into the galaxy system at a rate $\Phi$ and adds to the gas reservoir. Stars continuously form out of the reservoir at a rate that is assumed to be proportional to the mass of gas, characterized by a star-formation efficiency $\varepsilon$ or gas consumption timescale $\tau_{gas}$. A fraction of the stellar mass is immediately returned to the reservoir, along with newly produced metals. Finally, some gas may be expelled from the system, and possibly from the halo, by a wind $\Psi$ that is assumed to be proportional to the SFR. The mass of gas in the reservoir is free to vary and this regulates the star-formation. The picture on the right shows, in schematic form, the net flows through the system. The division of the incoming flow $\Phi$ into three streams is determined by $\varepsilon$, $\lambda$ and the sSFR, which are assumed to vary on timescales that are longer than the time the gas spends in the system, which is given by the gas consumption timescale $\tau_{gas}$.

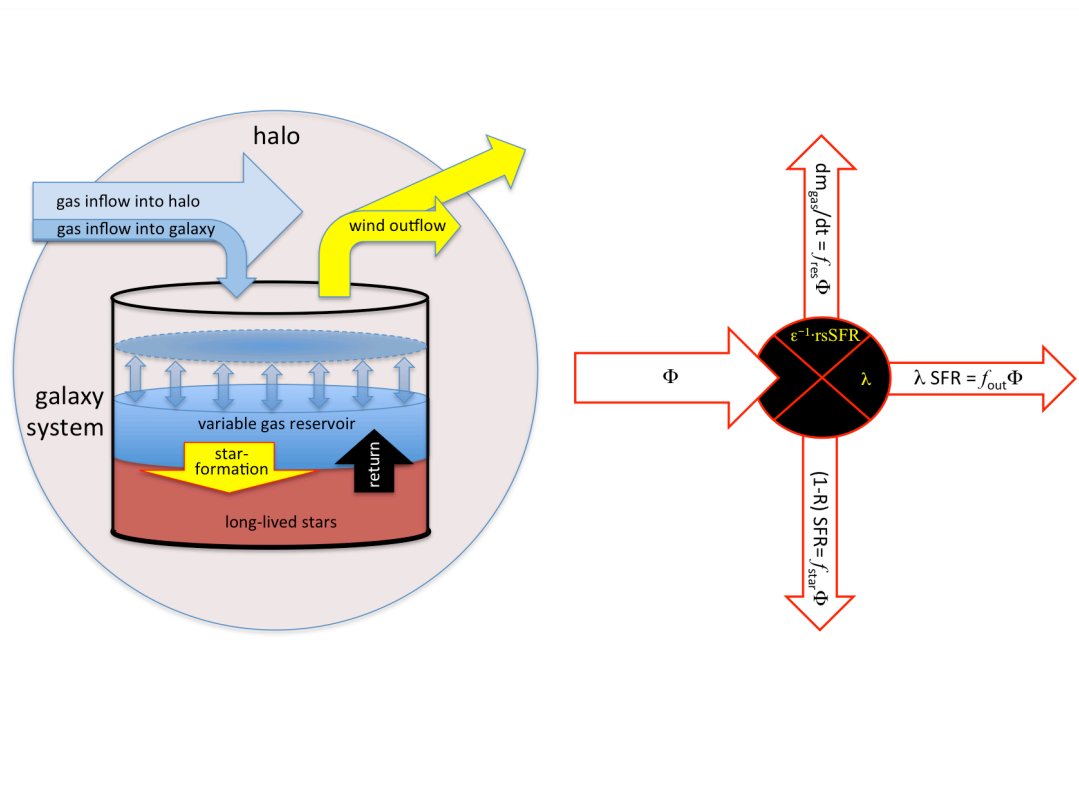

Figure 3: Heuristic examples of the behavior of the gas regulator under different illustrative circumstances. The upper two panels compare the input $sMIR_B$ (shown in black) and the resulting sSFR (in red) in four different situations - sudden increase in $sMIR_B$, sudden decrease in $sMIR_B$ (left-hand panels) and accelerating increase and decrease (right hand panels). In each case five different values of $\tau_{gas} = \varepsilon^{-1}$ are considered, varying logarithmically from 0.1 - 10 Gyr, i.e. from $\tau_{gas} = 10 \ sMIR_B^{-1}$ (dotted) to $\tau_{gas} = 0.1 \ sMIR_B^{-1}$ (dashed). The timescale for the response to sudden changes is the shorter of the $\tau_{gas}$ and the $sSFR^{-1}$ itself. The sSFR can track the accelerating increasing $sMIR_B$ with ease, but cannot track the decreasing case when $\tau_{gas}$ becomes longer than the timescale on which the $sMIR_B$ is changing. The lower two panels show the $m_{gas}/m_{star}$ ratio, $\mu$, that results in the same four histories of the $sMIR_B$ and the same five values of $\tau_{gas}$. See text for further discussion.

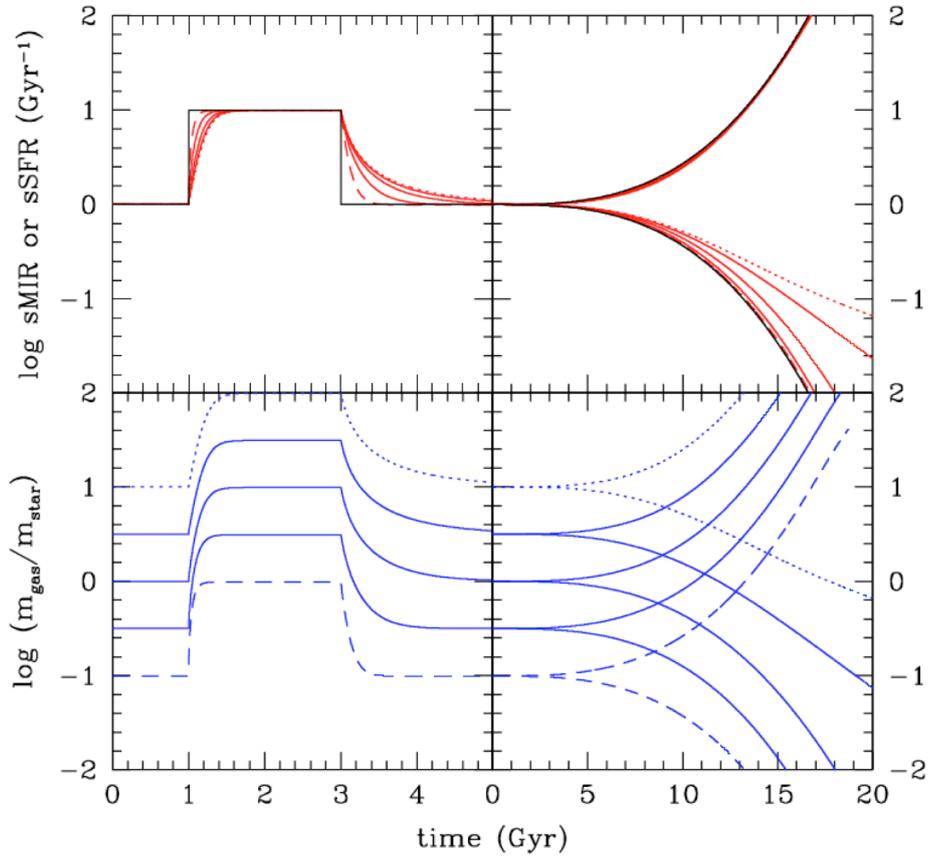

Figure 4: Different timescales relevant for the evolution of galaxies around $10^{10} M_\odot$. These are: the gas consumption timescale $\tau_{gas}$ from moecular observations (blue solid line, uncorrected for additional atomic gas), the $rsSFR^{-1}$ stellar mass increase timescale from observations (red solid line, with different possible behaviours at $z > 2$), the $sMIR_{DM}^{-1}$ dark matter mass increase timescale from simulations (black solid line), the timescales on which $rsSFR^{-1}$ and $sMIR_{DM}^{-1}$ are themselves changing $\tau_{rsSFR}$ and $\tau_{sMIRDM}$ (red and black dashed lines, respectively) and finally the Hubble timescale $\tau_H$ (solid magenta line) and the dynamical timescale of dark matter haloes $\tau_{dyn} \sim 0.1\tau_H$. The gas-regulation functions for as long as the gas consumption timescale $\tau_{gas}$ is (a) short compared with the timescale on which the inflow is changing, $\tau_{sMIRDM}$, which is satisfied at all $z$, and (b) short compared with the timescale on which the internal parameters of the regulator, $\epsilon$ and $\lambda$, are changing changing, which will be $rsSFR^{-1}$ if $\epsilon$ and $\lambda$ depend on stellar mass. Note how $\tau_{gas}$ is comparable to $rsSFR^{-1}$ at $z \sim 2$ and this may account for the change in the $rsSFR(z)$ behavior at higher redshifts as well as other changes in galaxian properties at earlier epochs.

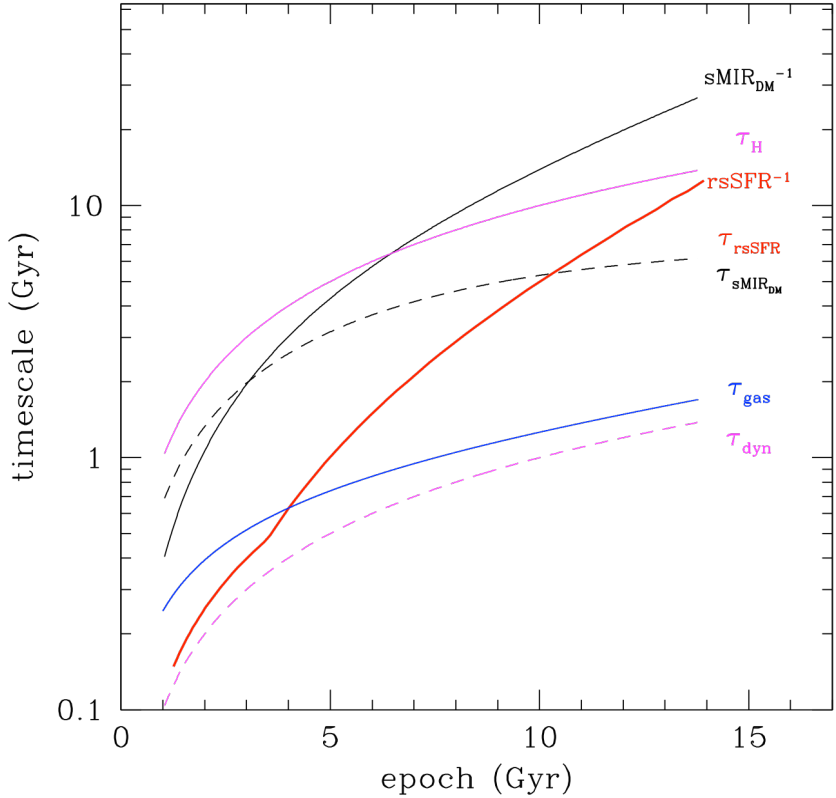

Figure 5:  Upper panel: The $Z(m_{star}, SFR)$ data for SDSS galaxies from Table 1 of Mannucci et al (2010, M+10).  Lower panel: residuals from the fit of Equation (40) to these data.

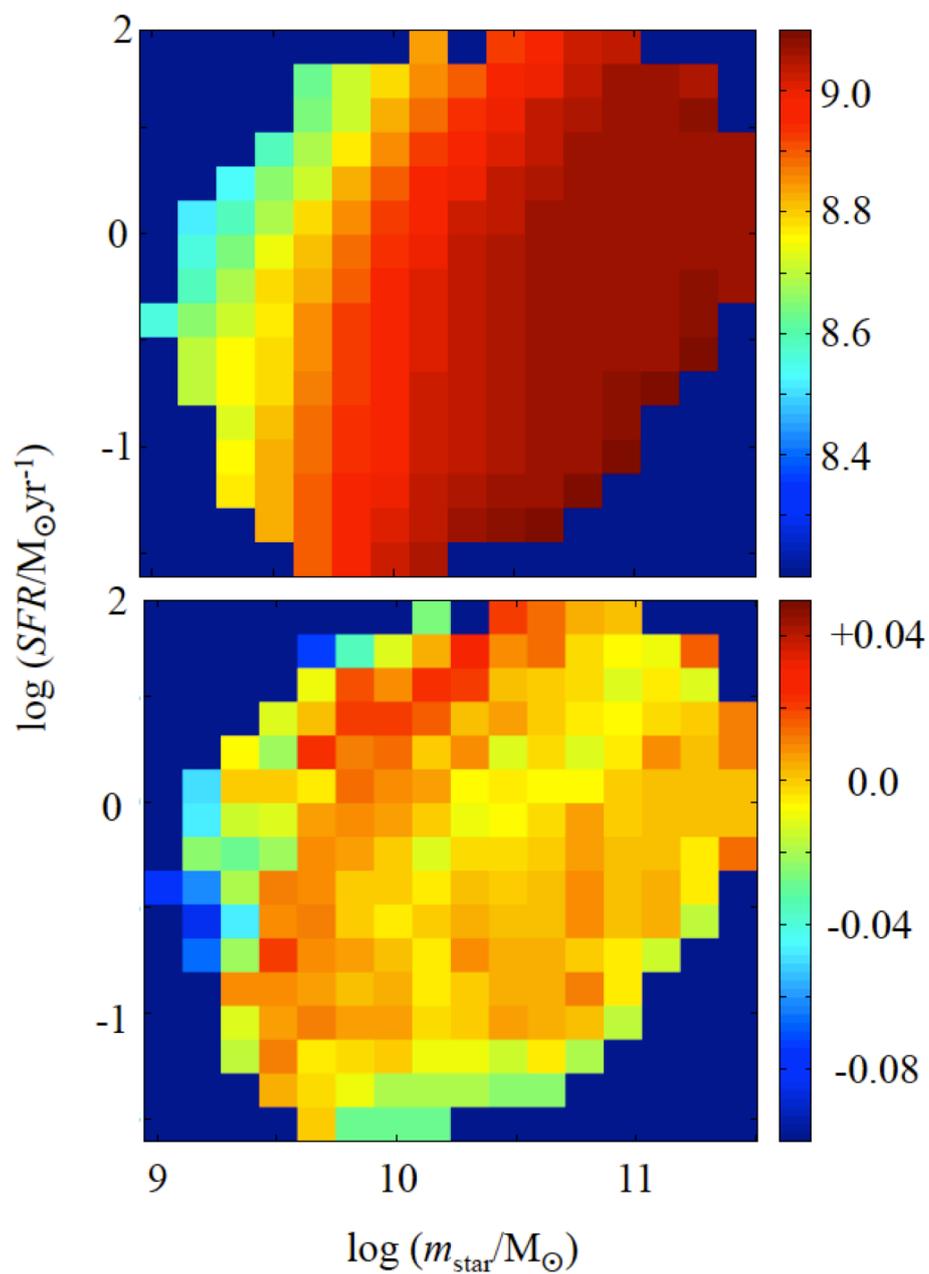

Figure 6: The destination of baryons that enter the galaxy system at $z \sim 2$ and at $z \sim 0$, derived from the expressions for $\varepsilon(m_{\text{star}})$ and $\lambda(m_{\text{star}})$ that are recovered by fitting the M+10 $Z(m_{\text{star}}, SFR)$ data and by fitting, with constraints, the T+04 $Z(m_{\text{star}})$ data. The three destinations for the baryons are shown as $f_{\text{star}}$ (red), $f_{\text{out}}$ (blue) and $f_{\text{res}}$ (green) with the shaded regions representing fits with $0.0 < Z_0/y < 0.1$. The formation of stars always dominates at very high masses and the ejection of material always dominates at low masses. At high $z$, the filling of the reservoir can dominate at intermediate masses on account of the high gas fractions in these systems. The greater curvature with mass in the low redshift Mannucci et al data produces a steeper mass-dependence on the mass-loading of the outflow. Note that $f_{\text{res}}$ is negative at low $z$, indicating that the reservoir is depleting, but this reversed flow is still small compared with the continuing infall rate, which has unit strength compared with the fractional quantities plotted here.

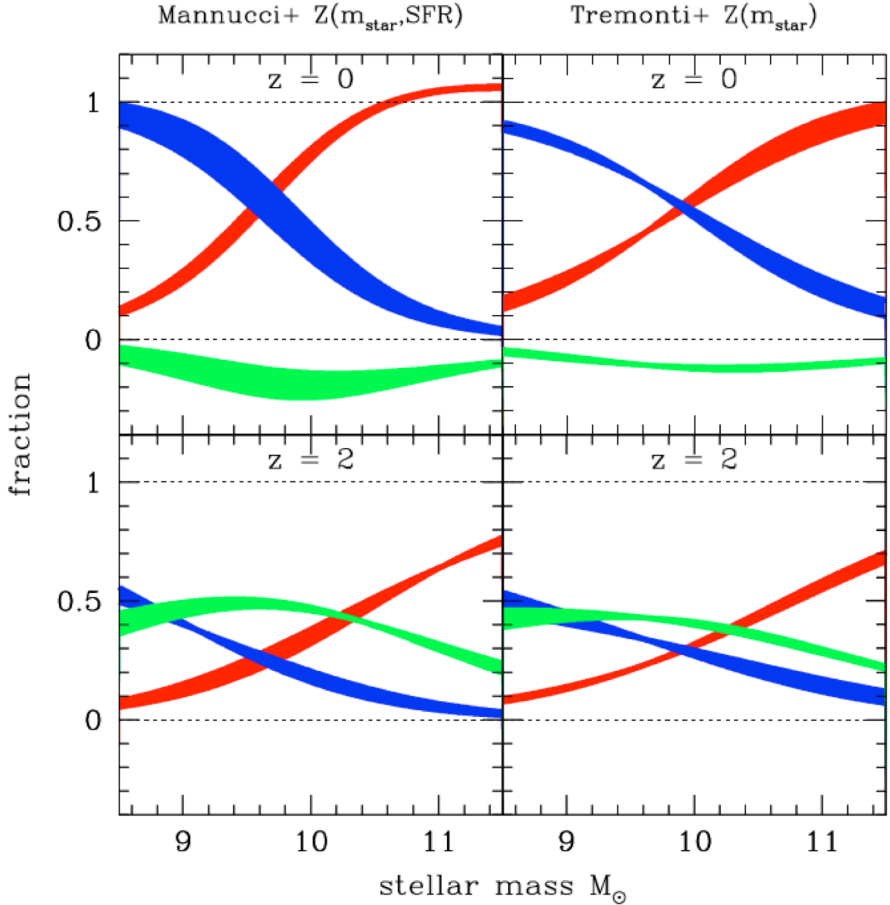

Figure 7: The mass-metallicity relations of Main Sequence star-forming galaxies at $z = 0$ (shown in red) and those predicted at higher redshifts $z = 1,2,3,4$ (continuous black lines, top to bottom) using the parameters from the fits of Equation (40) to the M+10 $Z(m_{star}, SFR)$ data (left hand panel) and the constrained fits to the T+04 $Z(m_{star})$ data (right hand panel). The blue points are from the NII/H$\alpha$ measurements of Erb et al (2006) transformed to the T+04 metallicity scheme using the formula in Kewley & Elliosn (2008) (blue dots, same in both plots). At $z = 0$ and at $z = 2$, the shaded areas show the range for fits with $0.0 < Z/y < 0.1$ (see Table 1). In deriving the predictions, the mean SFR-mass relations for Main Sequence galaxies based on Equation (2) are used. The outflow mass-loading $\lambda$ is taken to be constant with time, but the star-formation efficiency $\varepsilon$ is taken to increase as $(1+z)$, as indicated from observations (see text). While the low redshift curve must fit the data, by construction, the $z = 2$ curve is a straight prediction of the model, and successfully reproduces, at least qualitatively, the observed change in metallicity of Main Sequence galaxies. The dashed curves show the predicted metallicities at $z = 1,2,3,4$ (top to bottom) if $\varepsilon(m_{star})$ were to be held constant.

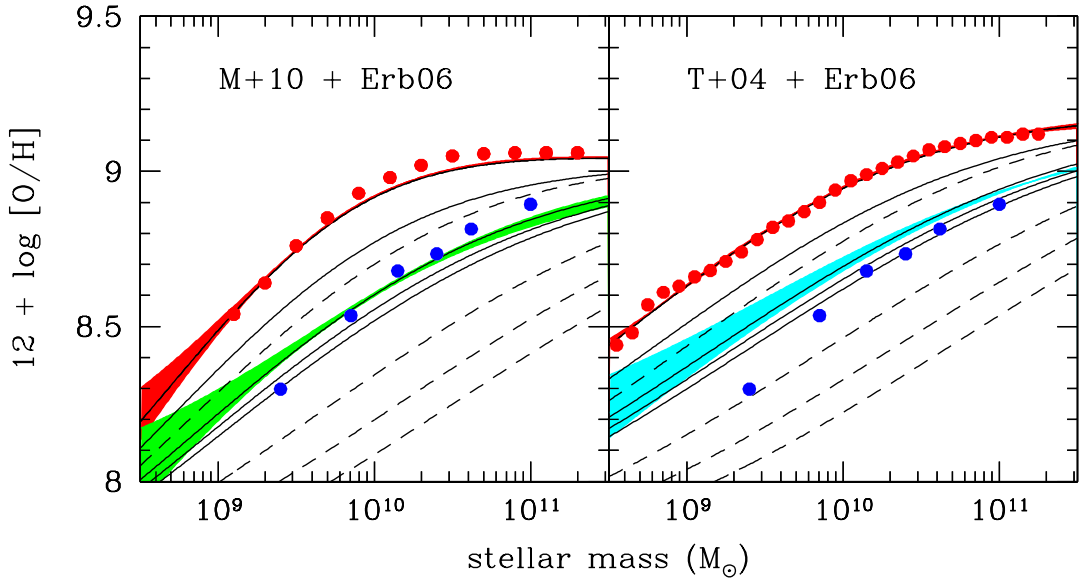

Figure 8. The ratio $m_{star}/m_{halo}$ as a function of stellar mass that is obtained from the simple numerical model described in the text, using the values of $\varepsilon(m_{star})$ and $\lambda(m_{star})$ for the $0 < Z_0/y < 0.1$ fits in Table 1, with $\varepsilon(m_{star})$ assumed to scale as $(1+z)$, as discussed in the text. The green curve is from the fits to the M+10 $Z(m_{star}, SFR)$ data, and the cyan one from the constrained fits to the Tremonti et al (2004) $Z(m_{star})$ data. The numerical model has been scaled to yield $m_{star}/m_{halo} = 0.03$ at the present-epoch for $m_{star} = 10^{10.5} M_\odot$, which requires a value of $f_{gal} = 0.4$. The left-hand panel shows the $m_{star}/m_{halo}$ ratio for galaxies with $m_{star} = 10^{10} M_\odot$ as a function of epoch, while the right hand panel shows the dependence on stellar mass at the present-epoch which reflects the slope of the mass-metallicity relation. The dashed black line shows the $m_{star}$-dependence ($\eta \sim 0.45$) required to reconcile the faint end slopes of the galaxy and halo mass functions (see text).

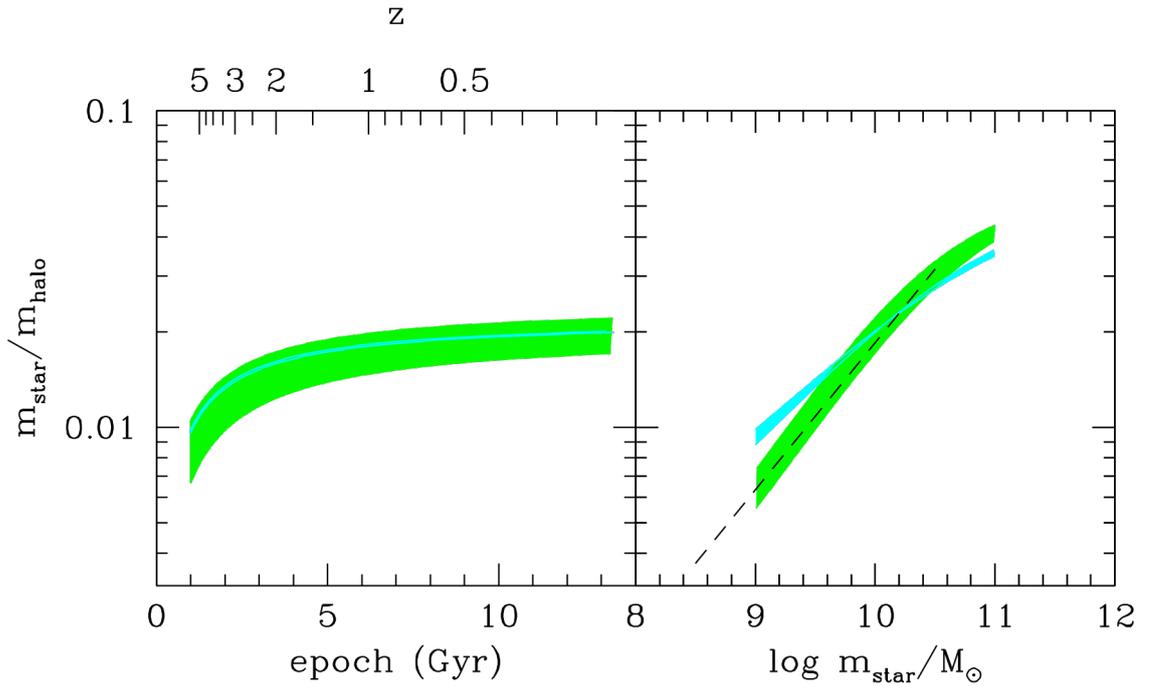

Figure 9. The observed rsSFR at $10^{10}M_\odot$ and the dark matter sSMIR$_{DM}$ at $10^{11.5}M_\odot$ (in red and black respectively) are carried over from Figure 1. The other curves show the predicted rsSFR that is output from the simple numerical model described in the text using the values of $\varepsilon(m_{star})$ and $\lambda(m_{star})$ for the $0 < Z_0/y < 0.1$ fits in Table 1 (green from the fits to the M+10 $Z(m_{star},SFR)$ data, cyan from the constrained fits to the Tremonti et al (2004) $Z(m_{star})$ data). The rsSFR is elevated because of the "catch-up" effect that arises because $f_{star}$ will increase as a galaxy grows, as shown in Figure 6). This boost, which is quite sensitive to the slope of the mass-metallicity relation through Equation (37), qualitatively reproduces the observed increase of the sSFR relative to the sMIR$_{DM}$. The apparent deficit at $z \sim 2$ could be related to the fact that the observed mass-metallicity relation may be steeper at high redshift than predicted by the simple model (see Figure 7). This would produce a larger $\eta$ and thus a larger boost, see text for discussion. The differential boost with stellar mass arises from the curvature of $f_{star}(m_{star})$ and qualitatively explains the reversed mass-dependence of the rsSFR compared with that of the sMIR$_{DM}$.

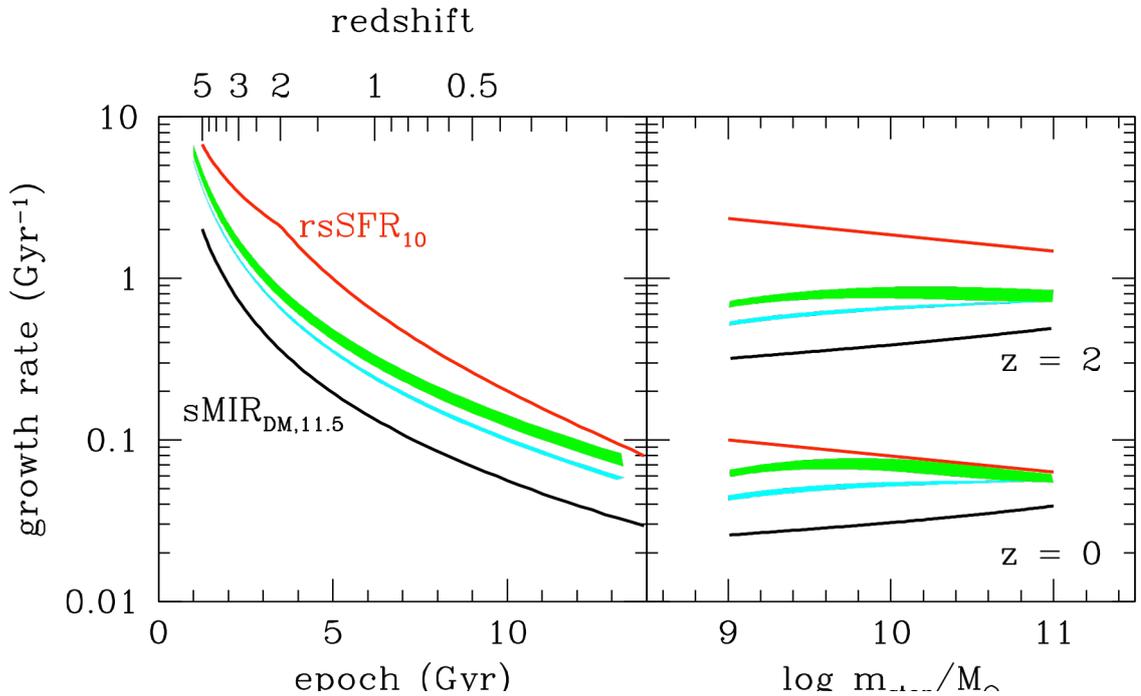

| $Z_0^{/} y$ | $\log y$[b] | $\lambda_{10}$ | $a$ | $\varepsilon_{10}^{-1}$ Gyr | $b$ |
|---|---|---|---|---|---|
| | | **Table 1: Fits of Equation (40) to observational data**[a] | | | |
| *Fits to Mannucci et al (2010) SDSS $Z(m_{star}, SFR)$ data* | | | | | |
| [0.00] | 9.02 | 0.25±0.02 | −0.81±0.03 | 2.4±0.2 | 0.28±0.03 |
| [0.03] | 9.00 | 0.29±0.03 | −0.79±0.03 | 2.8±0.2 | 0.32±0.03 |
| [0.10] | 8.98 | 0.40±0.04 | −0.77±0.03 | 3.8±0.3 | 0.41±0.04 |
| *Fits to Tremonti et al (2004) SDSS $Z(m_{star})$ relation*[c] | | | | | |
| [0.00] | 9.19 | 0.57±0.03 | −0.48±0.02 | [2.7] | [0.3] |
| [0.03] | 9.16 | 0.55±0.04 | −0.52±0.02 | [2.7] | [0.3] |
| [0.10] | 9.10 | 0.51±0.03 | −0.62±0.02 | [2.7] | [0.3] |

*Notes to Table:*

[a] *uncertainties on fitted parameters are formal uncertainties for each fit derived from the $\chi^2$ values relative to the best fit. The range of values in the Table gives a better indication of realistic uncertainties. Entries in [brackets] were imposed in the fits.*

[b] *expressed in units of 12–log(O/H)*

[c] *assumes a form for $\varepsilon$ and the Main Sequence SFR-$m_{star}$ relation from Equation (1) of Peng et al (2010)*